\documentclass[a4paper,11pt]{article}

\usepackage{jcappub}
\usepackage[utf8]{inputenc}
\usepackage{amsmath}
\usepackage{amssymb}
\usepackage{url}
\usepackage{graphicx}
\graphicspath{{./images/}} 
\usepackage{gensymb}
\usepackage{subfigure}
\usepackage{multirow, makecell}
\usepackage[capitalise]{cleveref}
\usepackage{tocstyle}
\usepackage{amssymb}
\usepackage{pifont}
\usepackage{siunitx}
\usepackage{hyperref}
\usetocstyle{standard}
\usepackage{tabularx}
\usepackage{booktabs}
\usepackage{ragged2e}
\usepackage{microtype}
\usepackage{braket}
\usepackage{comment}
\usepackage[normalem]{ulem}
\usepackage{fontawesome} 
\usepackage[dvipsnames]{xcolor}

\newcolumntype{Y}{>{\RaggedRight}X}
\newcolumntype{P}[1]{>{\RaggedRight}p{#1}}

\usepackage{placeins}

\usepackage{tikz}
\usetikzlibrary{automata,positioning,shadows}

\usepackage{xcolor}
\definecolor{c1}{HTML}{173f5f}
\definecolor{c2}{HTML}{20639b}
\definecolor{c3}{HTML}{3caea3}
\definecolor{c4}{HTML}{f6d55c}
\definecolor{c5}{HTML}{ed553b}

\newcommand{\charon}{\texttt{\raisebox{0.82\depth}{$\chi$}aro$\nu$}}
\newcommand{\pythia}{\texttt{P\footnotesize{YTHIA}\normalsize}}
\newcommand{\nusquids}{\texttt{$\nu$SQuIDS}}
\newcommand{\wimpsim}{\texttt{WimpSim}}
\newcommand{\geant}{\texttt{G\footnotesize{EANT}\normalsize4}}

\DeclareRobustCommand{\ammaG}{\text{\reflectbox{$\Gamma$}}}

\title{\charon{}: a tool for neutrino flux generation from WIMPs 
{\raisebox{1.00\depth}{\centering{\href{https://github.com/IceCubeOpenSource/charon}{\huge\color{BlueViolet}\faGithub}}}}
}

\author[a]{Qinrui Liu}
\author[a,b]{Jeffrey Lazar}
\author[b]{Carlos A. Arg{\"u}elles}
\author[c,d]{Ali Kheirandish}

\emailAdd{qliu246@wisc.edu}
\emailAdd{jeffrey.lazar@icecube.wisc.edu}
\emailAdd{carguelles@fas.harvard.edu}
\emailAdd{kheirandish@psu.edu}

\affiliation[a]{Department of Physics \& Wisconsin IceCube Particle Astrophysics Center, University of Wisconsin, Madison, WI 53706, USA}
\affiliation[b]{Department of Physics \& Laboratory for Particle Physics and Cosmology, Harvard University, Cambridge, MA 02138, USA}
\affiliation[c]{Department of Physics, The Pennsylvania State University, University Park, PA 16802, USA}
\affiliation[d]{Center for Multimessenger Astrophysics, Institute for Gravitation \& the Cosmos, The Pennsylvania State University, University Park, PA 16802, USA}

\abstract{
Indirect searches for signatures of corpuscular dark matter have been performed using all cosmic messengers: gamma rays, cosmic rays, and neutrinos. 
The search for dark matter from neutrinos is of particular importance since they are the only courier that can reach detectors from dark matter processes in dense environments, such as the core of the Sun or Earth, or from the edge of the observable Universe.
In this work, we introduce \charon{}, a software package that, in the spirit of its mythological Greek namesake $\chi \acute\alpha \rho \omega \nu $, bridges the dark sector and Standard Model by predicting neutrino fluxes from different celestial dark matter agglomerations.
The flux at the point of production is either computed internally by \charon{} or is taken from user supplied tables.
\charon{} then propagates this flux through vacuum or dense media and returns the expected neutrino yield at an observer's location.
In developing \charon{}, we have revisited and updated the production of neutrinos in dense media, updated the propagation of high-energy neutrinos, and studied the sources of uncertainty in neutrino transport.
This package is coupled to a new calculation that includes electroweak corrections resulting in the most up-to-date and complete repository of neutrino fluxes from dark matter decay and annihilation over the energy range of 1~GeV to 10~PeV coming from the Earth, the Sun, and the Galactic halo.
}

\begin{document}
\maketitle

\section{Introduction~\label{sec:intro}}

The Standard Model (SM) of particle physics---a theory with exceptional predictive power and confirmed by a plethora of measurements~\cite{PhysRevD.98.030001}---has failed to predict the existence of non-zero neutrino masses and provide a viable dark matter (DM) candidate.
In the search for beyond Standard Model Physics, neutrinos, DM, and the Sun are intimately connected.

Though the observation of non-zero neutrino masses has been confirmed using both anthropogenic and natural sources~\cite{Esteban:2018azc}, the first sign of neutrino flavor conversion was hinted by the mismatch between solar neutrino flux calculations~\cite{Bahcall:1998wm,Bahcall:2004pz} and data~\cite{Davis:1994jw}; see~\cite{Bahcall:2004cc,Lowe:2009sx} for historical summary and review.
Neutrinos, which interact only via the weak force, are the only cosmic messengers that can escape dense environments, such as the center of the Sun, unimpeded.
They yield information about their place of origin~\cite{Bahcall:2002jt} and their behaviour in extreme environments~\cite{Bahcall:1999ed}.
The existence of DM has been established by astrophysical and cosmological observations~\cite{Zwicky:1937, Kahn:1959, Freeman:1970, Rogstad:1972, Read:2014qva}.
In the standard $\Lambda$CDM model of cosmology, DM comprises about 85\% of the Universe's present matter content~\cite{Aghanim:2018eyx}; however, the nature and identity of DM are not currently known.
A number of DM candidates have been put forward, including massive compact halo objects (MACHOs), weakly interacting massive particles (WIMPs), axions, and sterile neutrinos.
Among these candidates, WIMPs are a class of elementary particles beyond the Standard Model which were thermally produced in the early Universe.
WIMPs are neutral and assumed to be stable over cosmological scales.
They interact gravitationally, and  are assumed to possess additional interactions at or below the weak scale; in fact, DM-neutrino interactions often appear in neutrino mass generation models~\cite{Ma:2006km,Boehm:2006mi,Farzan:2012sa,Escudero:2016tzx,Escudero:2016ksa,Alvey:2019jzx,Ballett:2019cqp,Ballett:2019pyw,Abdullahi:2020nyr}.
These so-called \textit{scotogenic} mass generation models have been proposed using sub-eV DM~\cite{Choi:2019zxy}, dark sectors below the electroweak scale~\cite{Abada:2014zra,Ballett:2019cqp}, and heavy DM~\cite{Bhattacharya:2018ljs,Chianese:2018dsz}.

WIMP DM can accumulate in the center of dense celestial bodies, such as the Sun, and annihilate into SM particles~\cite{Silk:1985ax,Hagelin:1986gv,Gaisser:1986ha,Srednicki:1986vj,Griest:1986yu}.
In the standard scenario, these particles can annihilate or decay to SM particles, yielding high-energy neutrinos that, unlike other products, can escape the Sun.
The detection of these neutrinos would provide a smoking-gun signature of DM since the only other source of high-energy solar neutrinos are those produced in cosmic-ray interactions in the solar corona, which have a small and well-predicted flux~\cite{Moskalenko:1993ke,Ingelman:1996mj,Hettlage:1999zr,Arguelles:2017eao,Edsjo:2017kjk,Ng:2017aur}, yet to be measured~\cite{Aartsen:2019avh}.
By looking for excesses of neutrinos coming from the direction of the Sun~\cite{Kamionkowski:1991nj, Bottino:1991dy, Halzen:1991kh, Gandhi:1993ce, Bottino:1994xp, Bergstrom:1996kp, Bergstrom:1998xh, Barger:2001ur, Bertin:2002ky, Hooper:2002gs, Bueno:2004dv, Cirelli:2005gh, Halzen:2005ar, Mena:2007ty, Lehnert:2007fv, Barger:2007xf, Barger:2007hj, Blennow:2007tw, Liu:2008kz, Hooper:2008cf, Wikstrom:2009kw, Nussinov:2009ft, Menon:2009qj, Buckley:2009kw, Zentner:2009is, Ellis:2009ka, Esmaili:2009ks, Ellis:2011af, Bell:2011sn, Kappl:2011kz, Agarwalla:2011yy, Chen:2011vda, Kundu:2011ek, Rott:2011fh, Das:2011yr, Kumar:2012uh, Bell:2012dk, Silverwood:2012tp, Blennow:2013pya, Arina:2013jya, Liang:2013dsa, Ibarra:2013eba, Albuquerque:2013xna, Baratella:2013fya, Guo:2013ypa, Ibarra:2014vya, Chen:2014oaa, Blumenthal:2014cwa, Catena:2015iea, Chen:2015uha, Belanger:2015hra, Heisig:2015ira, Danninger:2014xza, Blennow:2015hzp, Murase:2016nwx, Lopes:2016ezf, Baum:2016oow, Allahverdi:2016fvl, Rott:2012qb, Bernal:2012qh, Rott:2015nma, Rott:2016mzs}, neutrino observatories such as ANTARES, Baikal, Baksan, IceCube, and Super-Kamiokande, have set upper limits on the WIMP-nucleon scattering cross section~\cite{Adrian-Martinez:2016gti, Avrorin:2014swy, Boliev:2013ai, Aartsen:2016zhm, Desai:2004pq, Choi:2015ara}.
In the future, more stringent constraints can be set by Hyper-Kamiokande, Baikal-GVD, Km3Net, and, IceCube-Gen2 as they are expected to have improved sensitivities.

\begin{figure}[t!]
  \centering
  \includegraphics[width=\textwidth]{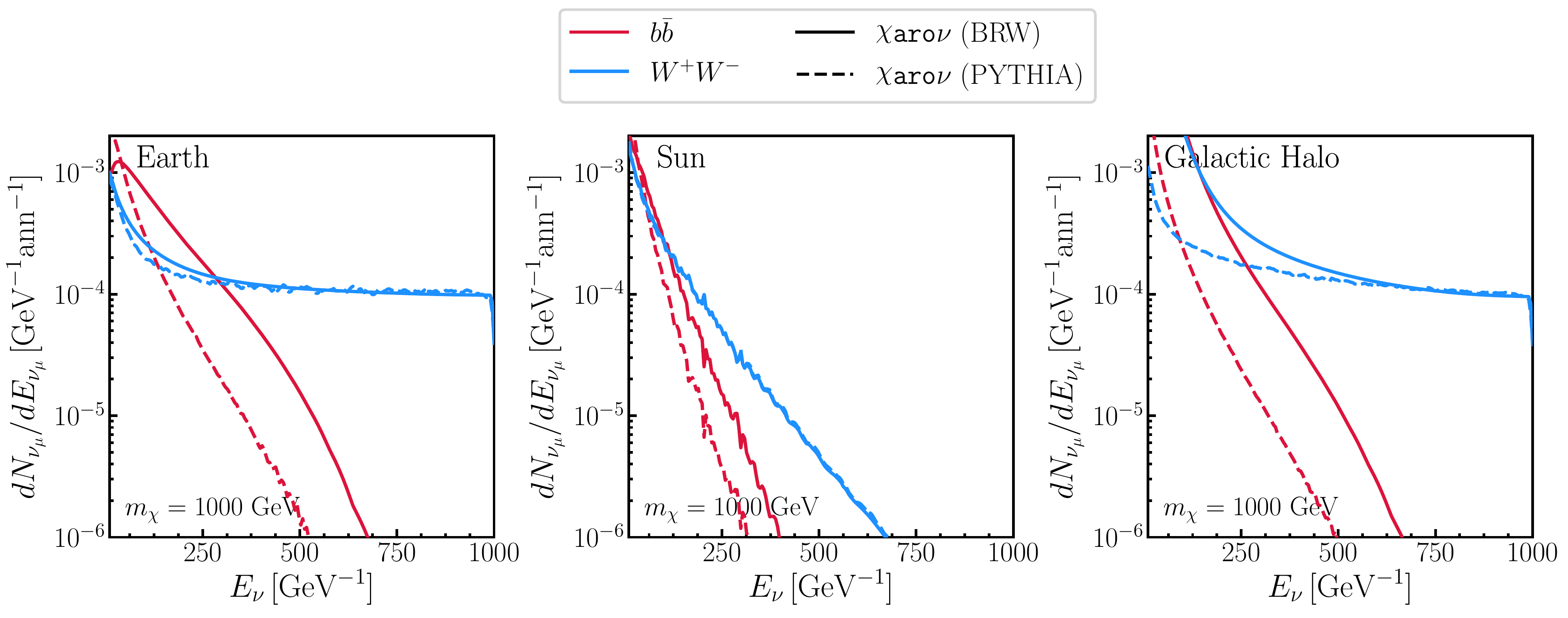}
  \caption{\textbf{\textit{\protect\charon{}'s predicted $\nu_{\mu}$ yield at Earth's surface for DM annihilation in the Earth, Sun, and the Galactic Halo.}}
  In this work we use the primary flux obtained from the recent BRW calculation~\cite{Bauer:2020jay}, which includes polarization of annihilation and decay products as well as a consistent treatment of the EW correction as our nominal flux.
  The dotted lines show our results using the primary flux from \pythia{}, which has a limited implementation of the EW correction.
  We show results for the normal mass ordering.}
  \label{fig:welcome_plot}
\end{figure}

Beyond the Sun, indirect searches look for evidence of WIMP annihilation or decay to SM particles in other regions that are expected to have large DM content, such as the Galactic center and Earth's core~\cite{Krauss:1985aaa,Freese:1985qw}.
Intensive efforts have also been made to detect such signatures from these objects, but so far only constraints have been placed by looking at the Earth~\cite{Albert:2016dsy,Aartsen:2016fep} and the Milky Way~\cite{Aartsen:2013dxa,Aartsen:2015xej,Avrorin:2015bct,Frankiewicz:2017trk,Iovine:2019rmd,Abe:2020sbr}; see~\cite{sergio_palomares_ruiz_2020_3959654} for a recent review.
Earth and Solar WIMP searches are complementary to direct detection strategies.
In the case of direct detection, one looks for energy deposited in a calorimeter by WIMPs scattering off nuclei.
The detection medium for these searches can be liquid noble gases such as argon or xenon~\cite{Aprile:2019xxb, Tan:2016diz, Akerib:2016vxi}, which provides the strongest constraint on the spin-independent WIMP-nucleon cross section; or an organic compound~\cite{Amole:2019fdf}, which is more sensitive to the spin-dependent WIMP-nucleon cross section.
This difference arises from the fact that axial-vector WIMP coupling to a nucleus is enhanced by having unpaired protons or neutrons~\cite{Jungman:1995df}.

In this work, we introduce a new tool, \charon{}~\cite{Virgil1902VA:b}\footnote{Pronounced Charon.}, to compute the neutrino fluxes from WIMP annihilations and decays in the Sun, the Earth, and the Galactic Halo.
For DM masses below the electroweak (EW) scale, the neutrino flux produced from DM annihilation or decay is computed using \pythia{}\texttt{-8.2}, while for masses above this scale a recent calculation by Bauer, Rodd, and Webber (BRW)~\cite{Bauer:2020jay}\footnote{The results of this calculation are publicly available \href{https://github.com/nickrodd/HDMSpectra}{here}.} is used.
The propagation of neutrinos from their production region to the detector is computed with the help of the neutrino propagation software \nusquids{}~\cite{arguelles:2015nu,Arguelles:2020hss}.
Our resulting fluxes below the EW scale are in good agreement with the \wimpsim{}~\cite{Blennow:2007tw} calculation, while above it we provide comparisons of our calculation with the existing calculations \wimpsim{} and the Poor Particle Physicist Cookbook (PPPC)~\cite{Ciafaloni:2010ti,Baratella:2013fya}.
Furthermore, we quantify the uncertainties in the transport of neutrinos that arises from neutrino cross sections and neutrino oscillation parameters.
For completeness, our package additionally contains a calculation of the WIMP capture rate in the Sun and Earth and a computation of the $J$-factor to be used for the Galactic contribution.
Additionally, our code also supports the secluded DM scenario~\cite{Pospelov:2007mp}, where DM annihilates to an unstable long-lived mediator which in turn decays to SM particles.
Figure~\ref{fig:welcome_plot}, shows our new calculation for the flux of DM annihilation in the Sun, Earth, and Galactic Halo with our estimated uncertainties due to neutrino transport.

The rest of this article is organized as follows: in Sec.~\ref{sec:detection}, we describe the theoretical background relevant for indirect DM searches.
In Sec.~\ref{sec:production}, we explain our procedure for generating neutrinos at the production site.
Sec.~\ref{sec:propagate} describes the procedure for propagating these fluxes from their birthplace to the detector.
In Sec.~\ref{sec:secluded}, we discuss the case in which there is a long-lived mediator that connects the dark sector to the SM.
Sec.~\ref{sec:compare}, compares our results with current widely used calculations \wimpsim{}~\cite{Blennow:2007tw} and PPPC~\cite{Cirelli:2010xx, Baratella:2013fya} as well as quantifies the total uncertainty associated with signal generation in neutrino indirect detection searches.
Finally, in Sec.~\ref{sec:conclusion} we conclude.

\section{Indirect Detection of Dark Matter\label{sec:detection}}

\begin{figure}[t!]
  \centering
  \includegraphics[width=0.85\textwidth]{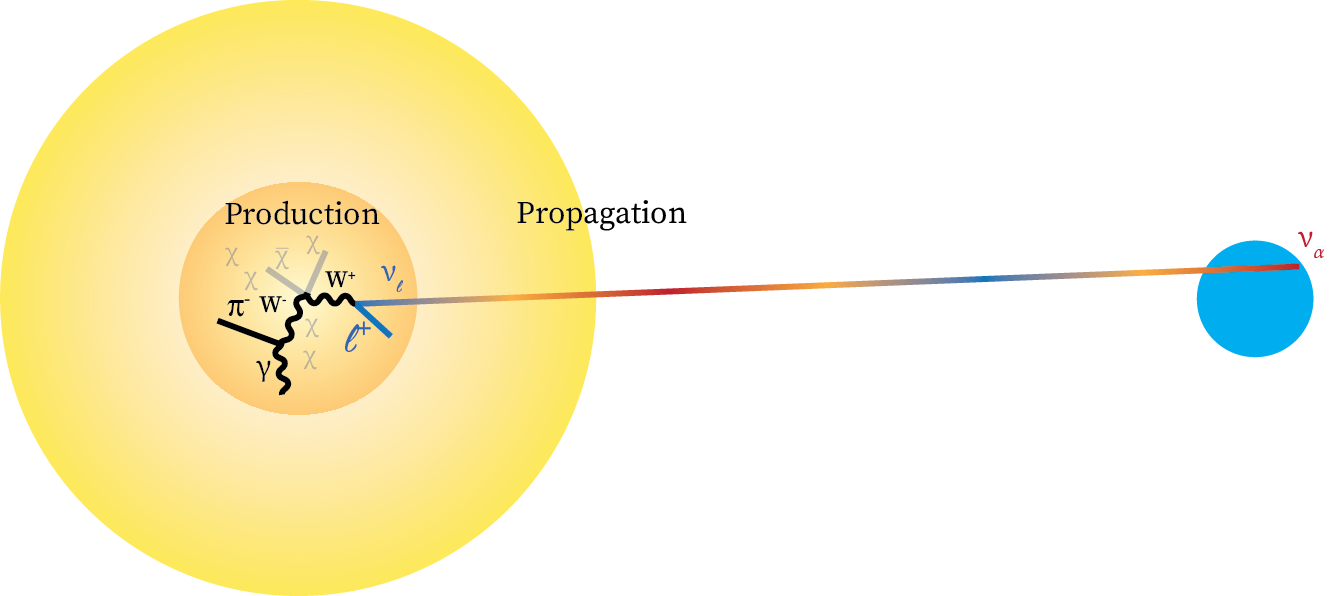}
  \caption{\textbf{\textit{Diagram depicting an example of WIMP annihilation to SM particles.}}
  These primary annihilation products, which in this illustration is a $W^{+}W^{-}$ pair, decay to other SM particles of which only neutrinos are able to leave the production region.
  The neutrino is indicated as a color-oscillating line to signify neutrino flavor change.} 
  \label{fig:standard_graphic}
\end{figure}
If DM is allowed to annihilate or decay to SM particles, experiments can detect these primary products of annihilation or decay or their secondaries as a signature of DM.
These attempts to find DM through such products are known as indirect detection searches.
In order to carry out these searches, one must know both where one expects large quantities of DM and the spectral shape of the signal.
In this section, we briefly discuss the production of neutrinos from these regions.

After an overdensity of DM has formed, DM can produce detectable signatures either via annihilation or decay to SM particles.
The spectral shape of the expected signal distribution depends on the primary channel, as well as any propagation effects, while the overall normalization is set by the rate of annihilation or decay.
Signals coming from the Galactic Halo are usually parametrized by the so-called $J$-factor, given by:
\begin{align}
    \label{jfactor}
    J_{\theta} =
    \begin{cases}
        \int_{\Delta\Omega} \int_{ \rm{l.o.s.}}\rho^{2}(\theta,\,l)dl\,d\Omega & ({\rm{annihilation}}),\\
        \\
        \int_{\Delta\Omega} \int_{\rm{ l.o.s.}}\rho(\theta,\,l)dl\,d\Omega &({\rm{decay}}),\\
    \end{cases}
\end{align}
where $\rho$ is the DM density profile, $\theta$ is the direction in the sky in a right-handed coordinate system centered at the Earth with one of its axes aligned with the direction to the Galactic Center (GC), and the integral is along the line-of-sight (l.o.s.). 
The l.o.s is related to the direction given by $\theta$ and the distance from the GC to the point of annihilation or decay, $r$, by $r^2 = r^2_\odot+ l^2-2lr_\odot\rm{cos}\theta$, where $l$ is the distance along the l.o.s.

After obtaining the $J$-factor, one can then compute the expected distribution of neutrinos from a given direction at the detector as
\begin{align}
    \Phi_{\nu,\rm{det}} = \frac{1}{4\pi}\frac{dN_{\nu}}{dE_\nu}J_\theta \times
    \begin{cases}
        \frac{\left<\sigma_{A} v\right>}{\kappa m^2_\chi} & ({\rm{annihilation}}),\\
        \\
        \frac{\Gamma_\chi}{m_\chi} & ({\rm{decay}}),\\
    \end{cases}
    \label{eq:phi_galactic}
\end{align}
where $d N_{\nu}/dE_{\nu}$ is the number of signal events per unit energy per annihilation or decay at the detector which differs from the flux at production by oscillation effects and energy redshift; $\Gamma_{\chi}$ is the rate at which the DM decays per DM particle; $\left<\sigma_{A}v\right>$ is the DM annihilation cross section averaged over the DM velocity distribution; and $\kappa$ is 2 for Majorana DM and 4 Dirac DM.
The DM velocity distribution is assumed to be a Maxwellian distribution in the Standard Halo Model, which is favored by recent simulations~\cite{Bozorgnia:2017brl}, though non-thermal components could be present due to accretion.
This averaging is necessary since in general the annihilation cross section is velocity-dependent, but its functional form is not known.
In fact, the above expression is only accurate for  velocity-independent dark matter interactions, since, in general, the velocity distribution is location dependent and cannot be factorized as in Eq.~\eqref{eq:phi_galactic}.
For generalization of the treatment above see~\cite{Ascasibar:2005rw,2009PhRvD..79h3525R,Ferrer:2013cla} for $J$-factor independent implementations, for treatments of $p$-wave annihilation~\cite{Campbell:2011kf,Diamanti:2013bia}, for Sommerfeld-enhanced cross section~\cite{Boddy:2017vpe,Lu:2017jrh,Bergstrom:2017ptx,Petac:2018gue}, for generalized $J$ factors~\cite{Boddy:2018ike, Boddy:2019qak}, and for a recent compendium and uncertainty estimation~\cite{Arguelles:2019ouk}.
Thus, for a given DM mass, and annihilation or decay channel searches looking at the GH are sensitive to the parameters $\left<\sigma_{A}v\right>$ or $\Gamma_{\chi}$, respectively.

Searches looking for DM signatures coming from the center of the Sun or Earth, on the other hand, are sensitive to the DM-nucleon scattering cross section.
To understand why this is, it is useful to examine the differential equation governing the number of DM particles in a source, given by
\begin{align}\label{eq:1}
    \frac{dN}{dt}=
    \begin{cases}
        C-AN^2-EN & (\rm{annihilation}),\\
        \\
        C-DN-EN & (\rm{decay}).
    \end{cases}
\end{align}
Here, the $C$ is the rate of WIMP accretion due to capture, $A$ is the thermally-averaged WIMP annihilation cross section per unit volume, and $D$ and $E$ are respectively the decay and evaporation rates per DM particle~\cite{Griest:1986yu, Jungman:1995df}.
For DM with masses above a few GeV, the rate of evaporation is expected to be negligible~\cite{Griest:1986yu, Gould:1987ju}.
Thus these equations are solved by
\begin{equation}
    N(t) = 
    \begin{cases}
        \sqrt{\frac{C}{A}}\tanh(\sqrt{C A}t) &(\rm{annihilation}),\\
        \\
        e^{-
        Dt} +\frac{C}{D} & (\rm{decay}).
    \end{cases}
\end{equation}
Whether or not the DM ensemble is in equilibrium depends on the capture and annihilation or decay rate; however, the DM ensemble is expected to be in equilibrium in the Sun~\cite{Catena:2016ckl}.
In the case of the Earth, on the other hand , this equilibrium condition is not expected to be satisfied~\cite{Catena:2016kro}.
When in equilibrium, \textit{i.e.} in the Sun, we can set the left hand side of Eq.~\eqref{eq:1} to zero, giving that the capture rate is given by
\begin{align}\label{eq:2}
    C =
    \begin{cases}
        AN^{2} \quad & (\rm{annihilation}),\\ 
        \\
        DN     \quad & (\rm{decay}).
    \end{cases}
\end{align}
This implies that the quantities experiments are sensitive to, \textit{i.e.} the annihilation and decay rates, are directly proportional to the capture rate at equilibrium.
By default, \charon{} calculates the capture rate originally given in~\cite{Gould:1991hx} with an up-to-date solar model~\cite{Vinyoles:2016djt}, but these can be changed by the user, using for example~\cite{Jungman:1995df,Halzen:2009vu,Athron:2018hpc,aaron_vincent_2020_3961678}.
When interactions are rare, the capture rate is linearly proportional to the DM-nucleon scattering cross section, $\sigma_{\chi N}$, which enables direct comparison to direct searches, which also probe $\sigma_{\chi N}$.
These two approaches though are complementary since they depend on different parts of the DM velocity distribution; the capture rate in the Sun is increased when DM is slower~\cite{Choi:2013eda}, while the direct detection scattering rate has the opposite effect.
This complementarity has been recently exploited to obtain velocity-independent constraints by combining results from the IceCube and PICO experiments~\cite{Amole:2019coq}.

From these sources, the expected flux at the detector can be expressed as
\begin{equation}\label{source_flux}
    \Phi_{\nu,\rm{det}} = \frac{{\ammaG}_{x}}{4\pi d^2}\frac{dN_\nu}{dE_\nu},
\end{equation}
where $\ammaG_{x}$ is the total rate of annihilation, $\ammaG_{\rm{ann}}$, or decay, $\ammaG_\chi$, from the source; and $d$ is the mean distance from the detector to the source.
The symbol $\ammaG_{\rm{dec}}$ should not be confused with the symbol $\Gamma_{\rm{dec}}$, which is the rate of decay per DM particle.

To compute the neutrino flux that is expected at a detector, the next steps are the computation of neutrino yield at the source and the propagation of those neutrinos from the source to the detector. A graphic illustrating these processes is shown in Fig.~\ref{fig:standard_graphic}. 

\section{Neutrino Yields from Annihilation and Decay: Algorithm Description\label{sec:production}} 

\begin{figure*}
  \centering
\tikzset{
diagonal fill/.style 2 args={fill=#2, draw=black, path picture={
\fill[#1, sharp corners] (path picture bounding box.south west) -|
                         (path picture bounding box.north east) -- cycle;}
                         },
reversed diagonal fill/.style 2 args={fill=#2, draw=black, path picture={
\fill[#1, sharp corners] (path picture bounding box.north west) |- 
                         (path picture bounding box.south east) -- cycle;}}
}
\tikzset{
pythianode/.style={
  rectangle,
  inner sep=3pt,
  text width=3cm,
  align=center,
  draw=black,
  fill=c3!40,
  line width=1
  }
}
\tikzset{
optnode/.style={
  circle,
  inner sep=3pt,
  text width=1.7cm,
  align=center,
  draw=black,
  fill=c4!40,
  line width=1
  }
}
\tikzset{
nsqnode/.style={
  rectangle,
  inner sep=3pt,
  text width=3cm,
  align=center,
  draw=black,
  fill=c5!40,
  line width=1
  }
}
\tikzset{
comment/.style={
  rectangle,
  inner sep=3pt,
  text width=2cm,
  align=center,
  draw=black!0,
  fill=c2!0,
  line width=1
  }
}

\resizebox{15cm}{!}{
\begin{tikzpicture}
\node (yes) at (4.5,1.4) [comment] {Yes};
\node (no) at (4.5,-1.4) [comment]{No};
\node (pdec) at (14.,-3.7) [comment] {Weight by $p_{\textrm{dec}}$};
\node (pdec) at (14.,-0.6) [comment] {Weight by $p_{\rm{int}}$};

\node (head) at (0,0) [diagonal fill={c3!40}{c2!40},
                       inner sep=3pt,
                       text width=3cm,
                       align=center,
                       line width=1
                      ]
      {Construct primary annihilation/decay products};
\node (issparse) at (3.7,0) [optnode] {Is the medium sparse?};
\node (decay) at (7.4,1.5) [pythianode] {Allow primary particle to decay to $\nu$};
\node (calcprobs) at (7.4,-1.6) [rectangle,
                                 inner sep=3pt,
                                 text width=3cm,
                                 align=center,
                                 draw=black,
                                 fill=c4!40,
                                 line width=1] 
      {Determine $p_{\rm{int}}$ and $p_{\rm{dec}}$};
\node (interact) at (11.7,-0.04) [pythianode] {Recursively simulate primary and secondary interactions and record $\nu$ yield.};
\node (weightdecay) at (11.7,-3.15) [pythianode] {Allow  particles to decay and collect $\nu$};
\node (prop) at (17.2,0) [nsqnode] {Collect all $\nu$ and propagate to detector.};

\node (extleg) at (-1.33,-1.7) [rectangle,
                           fill=c2!40,
                           draw=black,
                           line width=1,
                           minimum size=0.6cm
                          ] {};
\node (pythleg) at (-1.33,-2.4) [rectangle,
                           fill=c3!40,
                           draw=black,
                           line width=1,
                           minimum size=0.6cm
                          ] {};
\node (nsqleg) at (-1.33,-3.1) [rectangle,
                           fill=c5!40,
                           draw=black,
                           line width=1,
                           minimum size=0.6cm
                          ] {};
\node (extstr) at (0.6,-1.7) [rectangle,align=left] {BRW Calculation};
\node (extstr) at (0.2,-2.4) [rectangle,align=left] {PYTHIA 8.2};
\node (extstr) at (0,-3.1) [rectangle,align=left] {$\nu$SQuIDS};

\node (anchor) at (15,-0.0125) [rectangle] {};
\node (anchor2) at (14.5,1.5) [rectangle] {};
\node (anchor3) at (14.5,-3.15) [rectangle] {};
\node (anchor4) at (9.3,-3.15) [rectangle] {};
\node (anchor5) at (9.3,-0.06) [rectangle] {};
\node (anchor6) at (5,1.47) [rectangle] {};
\node (anchor7) at (5,-1.6) [rectangle] {};

\draw[->, line width=1.5] (head.east) -- (issparse.west);

\draw[-, line width=1.5] (issparse.north east) -- (anchor6.east);
\draw[->, line width=1.5] (anchor6.east) -- (decay.west);

\draw[-, line width=1.5] (issparse.south east) -- (anchor7.east);
\draw[->, line width=1.5] (anchor7.east) -- (calcprobs.west);

\draw[-, line width=1.5] (calcprobs.south east) -- (anchor4.east);
\draw[->, line width=1.5] (anchor4.east) -- (weightdecay.west);
\draw[-, line width=1.5] (calcprobs.north east) -- (anchor5.east);
\draw[->, line width=1.5] (anchor5.east) -- (interact.west);
\draw[-, line width=1.5] (interact.east) -- (anchor.east);
\draw[-, line width=1.5] (weightdecay.east) -- (anchor3.east);
\draw[-, line width=1.5] (anchor3.east) -- (anchor.east);
\draw[-, line width=1.5] (decay.east) -- (anchor2.east);
\draw[-, line width=1.5] (anchor2.east) -- (anchor.east);
\draw[->, line width=1.5] (anchor.east) -- (prop.west);

\end{tikzpicture}
}
  \caption{\textbf{\textit{Diagram of code structure.}}
  Flow chart depicting the major steps in the calculating flux from DM annihilation or decay. 
  The light yellow boxes indicate direct calculation or decision making; other colors indicate the main  program used in each step.
  }
  \label{fig:flow_chart}
\end{figure*}
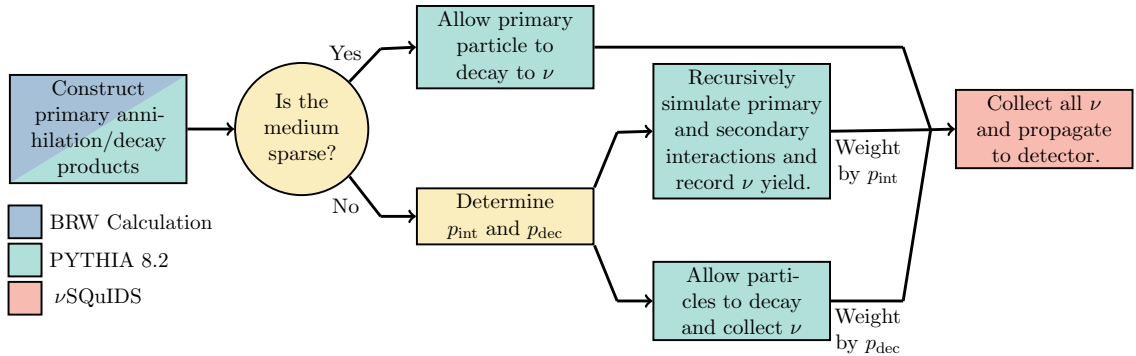

\begin{figure}[t!]
    \centering
  \includegraphics[width=0.9\textwidth]{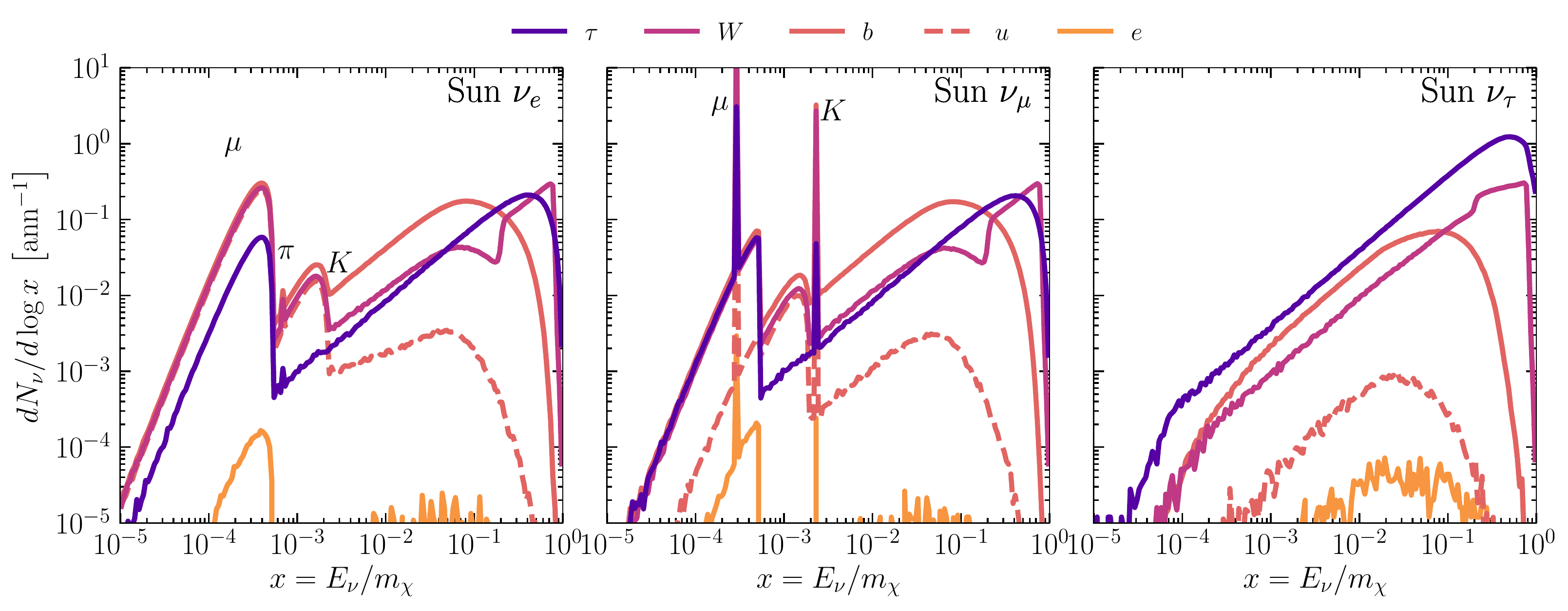}
  \includegraphics[width=0.9\textwidth]{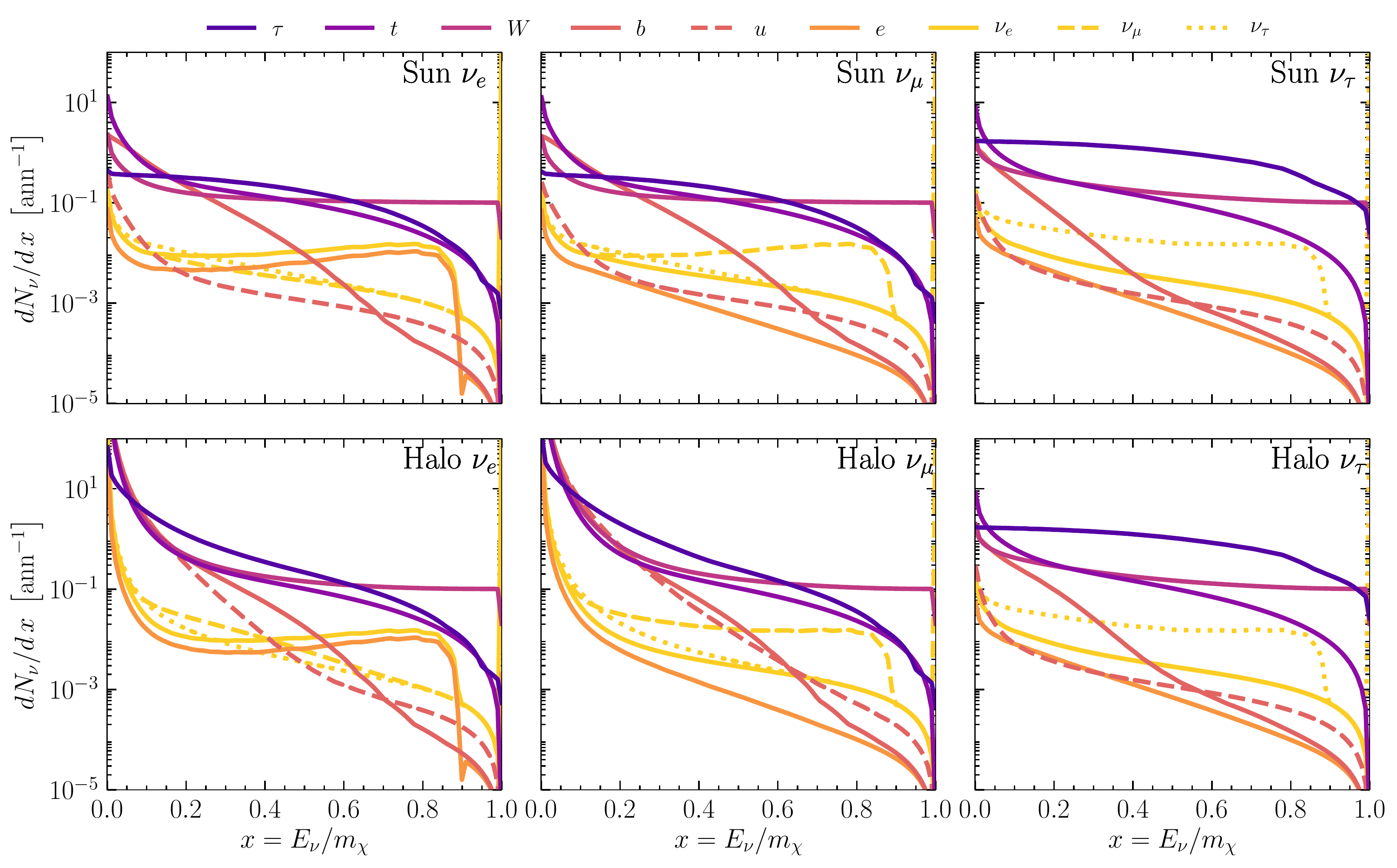}
  \caption{\textbf{\textit{Differential neutrino flux at production.}} We show spectra for several representative annihilation channels, indicated by different line colors.
  The top panel shows the spectra for $m_{\chi}=100$~GeV in the center of the Sun, generated without EW correction and with decays from stopped particles.
  Features related to $\mu$, $\pi$, and $K$ decays are indicated by their respective symbols.
  The two-row linear scale panel at the bottom shows spectra for a DM mass of $m_{\chi}=1000$~GeV with EW correction and without decays from stopped particles.
  Different polarization states are averaged over. 
  Due to EW corrections, direct neutrino channels no longer show a two-body decay distribution, but have a low-energy tail.
  The top row corresponds to the center of the Sun, while the bottom row is for the Galactic Halo. 
  The different columns indicate the neutrino flavor, from left to right: electron, muon, and tau flavored neutrinos. 
  For DM decay, $m_{\chi}$ should be multiplied by a factor of 2 and $x$ be replaced by $x = 2E_\nu/m_\chi$.
  Quarks not shown in this plot lie between the dashed and solid $u$ and $b$ lines, and differ only by a normalization.
  }
  \label{fig:production}
\end{figure}

We will now turn our discussion to the case of simulating the neutrino yield from DM annihilation and decay.
Here, we utilize a Monte Carlo approach to compute neutrino flux at production and integro-differential equation to propagate the neutrinos.
A high-level diagram of this process is shown in Fig.~\ref{fig:flow_chart} and the result of the algorithm is shown in Fig.~\ref{fig:production} for DM masses below and above the EW scale.

In our simulation, we assume that the parent DM particles are at rest.
This means that in the case of annihilation, each particle of the primary pair carries energy equal to the mass of the DM particle, $m_{\chi}$, and in the case of decay, each particle carries energy equal to $m_{\chi}/2$.
In both scenarios, the momenta of the primary pair are anti-aligned.
Therefore, the computation for both scenarios are identical up to the mass.
Therefore for conciseness, we will only focus on annihilation in the following discussion.

Since the interaction of DM with the SM is unknown, we take the pragmatic approach of computing the neutrino spectra for DM annihilation to $q\bar{q}$, $gg$, $W^{+}W^{-}$, $Z^{0}Z^{0}$, $HH$, $l^{+}l^{-}$, and direct neutrino channels $\nu_{\alpha}\bar{\nu}_{\alpha}$; the latter channel is relevant in the case of Kaluza-Klein DM~\cite{Hooper:2002gs}. Here, $q\bar{q}$ are the six quarks, $l^{+}l^{-}$ are the three charged leptons, and $\nu_{\alpha}\bar{\nu}_{\alpha}$ are the three neutrino flavors.
The neutrino yield depends on the interplay of two variables: the primary particles produced and the environment in which production takes place.
In sparse environments, those where the interaction lengths of particles are much longer than decay lengths, neutrinos can be formed as the primary decay or annihilation products, or as secondary products produced after the primary particle hadronizes and showers.
In dense environments, on the other hand, this condition is not satisfied; the competition between interactions of the produced particles with matter and their decay process modifies the final neutrino spectrum.
Therefore, one must compute the secondary particles' interaction lengths in order to precisely predict the neutrino fluxes.

In this work, we compute the fluxes at production using \pythia\texttt{~8.2} with some modifications, since a vacuum environment is assumed by default in \pythia{}.
For WIMP masses above 500~GeV, we use the BRW calculation of the initial flux.

After accounting for hadronization and final states radiation of the primary particles, we simulate the interaction of the products with their environment.
This involves determining whether or not the environment is sparse as previously defined, which in general depends on the particle.
For all particles considered in this paper, the Galactic Halo is a sparse environment, and as such all particles produced there are allowed to decay without interacting.
In the case of the core of the Sun or Earth, an evaluation of particle interactions is needed. 
Here, we discuss energy losses of hadrons, created from hadronization of quarks, and charged leptons in matter.
In the following text, we will enumerate our treatment of particle interactions by cases.

\textbf{Hadrons:}
Except the top quark, which will immediately decay, all quarks will promptly hadronize.
The resulting hadrons will interact with the matter in the production environment.
The mean interaction time, $\tau_{\rm{int}}$, for a given hadron is given by
\begin{equation}
    \tau_{\rm{int}} = \left(n\sigma v\right)^{-1}
                    \approx 5.5\times 10^{-8}\;s\;\times\left(\frac{\rm{g}/\rm{cm}^{3}}{\rho}\right)\left(\frac{\rm{mb}}{\sigma}\right)\beta^{-1},
\label{eq:tint}
\end{equation}
where $\rho$ is the density of the medium, $\sigma$ is the hadron-nucleon scattering cross section, and $\beta$ is the velocity of the hadron in natural units.

When the lifetime of the hadron is very short compared to the mean interaction time, the hadron is allowed to decay freely since energy losses due to interactions will be negligible.
When the interaction and decay times are comparable, which is the case of hadrons that contain bottom ($b$) or charm ($c$)  valence quarks, we weight our simulation according to the rate of these two processes.
To carry out this weighting, we assume that hadrons containing $b$ or $c$ quarks have a similar hadron-nucleon cross sections, which are approximately, $\sigma_{b/c-\mathrm{meson}}\sim14~\mathrm{mb}$ for mesons and $\sigma_{b/c-\mathrm{baryon}}\sim24~\mathrm{mb}$ for baryons~\cite{Edsjo:1997hp}. 

When a hadron containing $b$ or $c$ valence quarks is created in the simulation, we assume there is a probability of interacting and a probability of decaying given by
\begin{equation}
p_{\rm{int}}=\frac{\gamma \tau_{\rm{dec}}}{\tau_{\rm{int}}+\gamma \tau_{\rm{dec}}}
~{\rm and}~
 p_{\rm{dec}}=\frac{\tau_{\rm{int}}}{\tau_{\rm{int}}+\gamma \tau_{\rm{dec}}},
\end{equation}
respectively, where $\tau_{\rm{dec}}$ is the rest-frame lifetime of the hadron and $\tau_{\rm{int}}$ is the mean time between interactions.
Both interaction and decay are then simulated.
In the case of decay, all products are weighted by $p_{\rm{dec}}$.
In the case of interaction, one the other hand, the hadron is given a new energy $E'$ and weighted by $p_{\rm{int}}$, and this process is recursively carried out until the hadron has lost all kinetic energy.
In this procedure, low-energy neutrinos produced in hadronic showers initiated by the aforementioned hadron are not tracked, and thus we underestimate the low-energy flux; see~\cite{Bernal:2012qh} for a detailed low-energy calculation.
In this work, we set $E' = \left<Z_{x}\right>E_{0}$ where is $\left<Z_{x}\right>$ is the average fractional energy transfer, and $E_{0}$ is the energy of the hadron before interaction.
We follow the estimates of~\cite{Ritz:1987mh} and set $\left<Z_{b}\right>=0.7$ and $\left<Z_{c}\right>=0.6\frac{m_{c}}{m_{\mathrm{hadron}}}$.
\charon{} and \wimpsim{} use this energy loss estimation method, while PPPC samples from an energy loss distribution; we have checked that these two approaches yield comparable results.

Using Eq.~\eqref{eq:tint} to estimate the ratio of $\gamma\tau_{\rm{dec}}$ to $\tau_{\rm{int}}$ for hadrons composed only of up, down, and strange quarks gives a number of order $10^{4}$ and $10^{3}$ in the Sun and Earth respectively for $m_{\chi} = 1~{\rm GeV}$.
These ratios grow to $10^{8}$ and $10^{7}$ for $m_{\chi} = 1~{\rm TeV}$.
Thus, one can safely assume that such hadrons, namely $\pi^{\pm}$, $K^{\pm}$, $K_{L}^{0}$, and neutrons are fully stopped on account of rapid interactions before decaying.

\textbf{Charged leptons:}
The rate at which charged leptons lose energy in a plasma, such as the center of the Sun, is given in~\cite{Jackson:100964}.
We find that the rate of energy loss, $\frac{dE}{dt}$, is larger than $10^{9}~\mathrm{GeV} / \mathrm{s}$.
This gives that the stopping time is bounded by $\tau_{\mathrm{stop}} = m_{\chi}/\left(dE/dt\right)$.
Computing the ratio of  $\tau_{\mathrm{stop}}$ to $\gamma \tau_{\mathrm{dec}}$ for tauons ($\tau$) and muons ($\mu$) leads to the conclusion that in the relevant energies, the $\tau$ will promptly decay, but most $\mu$ will stop before decaying.

In the Earth, charged leptons lose energy through ionization, bremsstrahlung, photo-nuclear interaction, and electron pair production, which we estimate using the Bethe-Bloch formula~\cite{PhysRevD.98.030001}.
The interaction length of $\tau$ in Earth is significantly larger than the decay length up to $\tau$ energies of $10^{9}~\rm{GeV}$~\cite{Koehne:2013gpa}.
Thus for the DM mass range under discussion in this work, one can assume the $\tau$ decays at production without interacting.
Muons are longer lived, and at $\mu$ energies above a few GeV the interaction length is much shorter than the decay length.
As in the case of the Sun, they can be assumed to come to rest before decaying.

\textbf{Long-lived particles:} Long-lived hadrons, $\pi^\pm$, $K^\pm$, $K^0_L$, and neutrons, are either fully absorbed by matter or decay after being stopped.
Among them, $\pi^-$ and $K^-$ are captured by matter and would form atom-like systems, which prohibits them from further decays~\cite{Ponomarev:1973ya}.
Due to different interactions of $K^0$ and $\bar{K}^0$ with nucleons, which causes quantum coherence loss, $K^0_S$ are regenerated from $K^0_L$ continuously and followed by hadronic decays~\cite{Good:1957zza}.
Stopped $\mu^\pm$, $\pi^+$, and $K^+$ continue to decay at rest.
Neutrinos from decay of stopped particles are produced at energies below 1~GeV, thus are not of great importance for current indirect searches.
\charon{} provides an option to compute these low-energy neutrinos using \pythia{}.
These neutrinos have also been considered in previous calculations, \textit{e.g.} they are included in PPPC using \geant{}~\cite{Baratella:2013fya}, a technique first developed in~\cite{Bernal:2012qh}; see also~\cite{Rott:2012qb,Rott:2015nma,Rott:2016mzs}.

\textbf{Polarizations and EW corrections:} 
For DM masses above the EW scale, the neutrino flux generation in \charon{} is coupled to a new computation of the EW correction which also accounts for polarization of annihilation and decay states and the evolution of the polarized particles to the EW scale~\cite{Bauer:2020jay}. 
As EW interactions are not fully implemented in \pythia{}, this calculation incorporates a more complete consideration such as the missing triple gauge couplings in the EW sector.
Polarization and EW correction effects are also implemented in PPPC, which takes a different approach by augmenting leading order EW correction term~\cite{Ciafaloni:2010ti}.
Other works have studied the effects of polarization of annihilation or decay products without EW correction in~\cite{Garcia-Cely:2016pse,Niblaeus:2019ldk}.
We find that the inclusion of this new estimation of EW corrections and polarization gives rise to different spectra, which may be softer or harder depending on the channel and DM mass.

\section{Neutrino Transport with \nusquids{}\label{sec:propagate}}

As in Sec.~\ref{sec:production}, we discuss two cases of interest: neutrino-opaque and neutrino-transparent media.
To distinguish between these scenarios, we define a neutrino-opaque environment as those where the neutrino interaction length is less than the distance the neutrino needs to travel to exit the source and neutrino-transparent environments otherwise.
The neutrino interaction length is an energy dependent quantity given by
\begin{equation}
    \lambda^{\rm{int}}(E_\nu) = \left[n_N \sigma^{\rm{tot}}(E_\nu)\right]^{-1} = \left[n_N \sigma^{CC}(E_\nu) + n_N \sigma^{NC}(E_\nu)\right]^{-1},
\end{equation}
where $n_N$ is the number density of nucleons, $\sigma^{CC}$ is the neutrino-nucleon charged-current cross section, and $\sigma^{NC}$ the neutral-current cross section; the neutrino-electron cross section is a subdominant contribution to the interaction rate and can be neglected in most applications~\cite{Gandhi:1998ri}.
These latter terms define, respectively, the charged- and neutral-current interaction lengths~\cite{Gandhi:1998ri}.

For electron and muon neutrinos with energies above 100~GeV~\cite{Formaggio:2013kya}, the interaction length is essentially identical, and thus the interaction rate will be the same for both flavors of neutrinos.
On the other hand, the $\nu_{\tau}$ charged-current cross section differs significantly from the $\nu_{e}$ and $\nu_{\mu}$ charged-current cross sections due to phase-space suppression, which results in a 30\% difference at 100~GeV~\cite{Kretzer:2002fr,Paschos:2001np,Hagiwara:2003di}; see~\cite{Conrad:2010mh} for a discussion of this in neutrino oscillation experiments.
This means that below this energy, environments will be less opaque to tau neutrinos than to neutrinos of other flavors.
Furthermore, the neutrino yield is flavor-dependent because the competition between interaction and decay differs for each flavor of charged lepton.
We do not consider neutrinos from electrons since they do not decay, and, in this work, we do not include neutrinos produced by showers induced by electrons.
Muons can decay, but since they are relatively long-lived, they will interact with a medium as discussed in Sec.~\ref{sec:production}, losing energy and reducing the yield of high-energy neutrinos.
Tauons, on the other hand, are short-lived, and will decay quickly enough to give a significant fraction of their energy to the produced neutrino in a process known as tau-neutrino regeneration~\cite{Halzen:1998be}.
To be concrete, for energies less than 10~PeV and densities of $\mathcal{O}\left(\rm{g}/\rm{cm}^3\right)$, the tauon will decay promptly to a hard neutrino.

After accounting for the flavor dependence of secondary neutrino yield due to both the yield of neutrinos per charged-current interaction and the interaction rate itself, one must consider the effect of neutrino flavor change, commonly called neutrino oscillations~\cite{Giunti:2007ry}.
This phenomena is due to the fact that the mass and flavor bases are misaligned.
These two bases are related by a unitary transformation known as the Pontecorvo-Maki-Nakagawa-Sakata (PMNS) or neutral-lepton-mixing matrix, $U$.
This three-by-three, complex-valued matrix can be parameterized as a product of two real and one complex rotations, namely $U = R_{23}\bar R_{13}R_{12}$~\cite{PhysRevD.98.030001} where $R_{ij}$ is a real rotation and $\bar R_{13}$ a complex rotation.
The real angles involved in this parameterization have been measured to percentage levels~\cite{Esteban:2018azc} and there is hint of a non-zero complex phase~\cite{Abe:2019vii}.
The elements of this matrix, which relate the flavor and mass states, dictate the flavor transition amplitude, while the characteristic length over which transitions occur depends on the mass-squared differences between the mass eigenvalues.
This oscillation length is given by
\begin{equation}
    \lambda^{\rm{osc}, i} = 4 \pi \frac{E_\nu}{\Delta m^{2}_{i1}},
    \label{eq:osc_length}
\end{equation}
where $\Delta m^2_{i1}$, with $i = 2~{\rm or}~3$, are the two neutrino mass-squared-differences which have been measured from terrestrial and solar neutrino experiments and are known to the few-percent level~\cite{Esteban:2018azc}.

In this work, we use the \nusquids{}~\cite{arguelles:2015nu,Arguelles:2020hss} software package to consistently account for the above effects in neutrino transport.
This package represents the initial flavor in the density matrix formalism, $\rho_\alpha = \ket{\nu_\alpha}\bra{\nu_\alpha}$, the evolution of which is governed by
\begin{equation}\label{eq:nu_evol}
\frac{\partial \rho}{\partial r}=-i\left[H, \rho\right]-\left\{\Gamma, \rho_{i}\right\}+\int_E^{\infty} F(\rho,\bar{\rho},r,E',E)dE',
\end{equation}
where $H$ is the full neutrino Hamiltonian, including both the neutrino kinetic terms and matter potentials; $\Gamma$ is the interaction rate; and $F$ is a functional which encodes the cascading down of neutrinos due to charged and neutral current processes, see~\cite{Arguelles:2020hss} for details.
The terms in this equation depend on the environment traversed by the neutrino, as well as the PMNS matrix and neutrino cross sections.
For our nominal results and plots shown in this work we set the oscillation parameters to the best-fit values for normal ordering from NuFit-4.1~\cite{Esteban:2018azc, nufit:2018abc} and model the neutrino nucleon cross section using the \texttt{nuSIGMA} software package~\cite{edsjo2007calculation}.

After the neutrinos exit their production region---in this work the Sun center, the Earth center or the Galactic Halo ---they must be propagated to the detector location at Earth.
In this voyage, we need to account for neutrino oscillations in vacuum and propagation inside Earth if the flux traverses significant matter, which imply that the spectrum for  $\nu_{\alpha}$ at the detector can be written as
\begin{equation}
\frac{dN^{\oplus}_{\nu_\alpha}}{d\Omega dE_{\nu_\alpha} dt} = f(\Omega)\sum_{\beta}^{3}\frac{dN^\mathrm{prod}_{\nu_\beta}}{dE_{\nu_\beta}}\bar P_{\nu_\beta\rightarrow \nu_\alpha}(d, E_\nu),
\label{eq:spectra}
\end{equation}
where $dN^\mathrm{prod}_{\nu_\beta}/dE_{\nu_\beta}$ is the neutrino yield per annihilation or decay computed in Sec.~\ref{sec:production}; $\bar P_{\nu_\beta\rightarrow \nu_\alpha}(d, E_\nu)$ is the production-region-averaged probability of $\nu_\beta$ oscillating to $\nu_\alpha$ at distance $d$ with energy $E_\nu$; and $f(\Omega)$ is a function which encodes information about the rate of neutrino production and the geometry of the source.
The rate $\ammaG$ is process- and location-dependent and the specifics of this dependence are discussed in Sec.~\ref{sec:detection}.

\begin{figure}[t!]
\centering
  \includegraphics[width=0.9\textwidth]{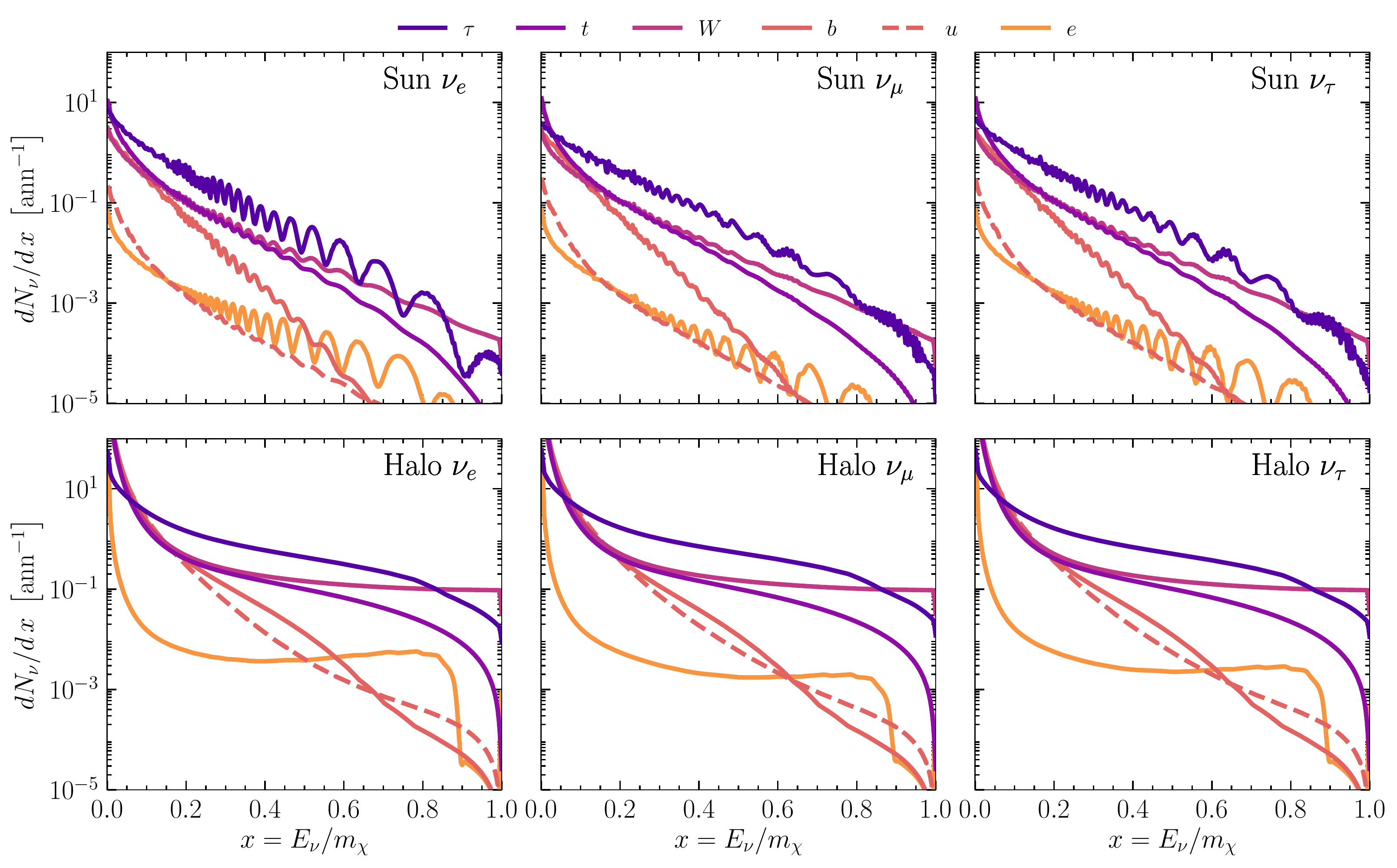}
  \caption{\textbf{\textit{Neutrino flux from DM annihilation at Earth's surface.}}
  Colors and line styles have the same meaning as in Fig.~\ref{fig:production}.
  Results in the top panel are computed with a zenith angle of $60^\circ$ while those in the bottom panel  are computed with a zenith angle of $180^\circ$.
  The DM mass have been set to 1~TeV for annihilation; for decay it should be read as 2~TeV and $x$ should be replaced by $x = 2E_\nu/m_\chi$.} 
  \label{fig:detector}
\end{figure}

In what follows, we give additional details relevant for the propagation in the three production environments considered in this work.
The result of these processes is summarized in Fig.~\ref{fig:detector}.

\subsection{Galactic Flux Propagation}

The distance between the Galactic Center and the Earth is sufficiently large that all current- and next-generation detectors do not have sufficient energy resolution to resolve individual oscillations.
Thus, the energy and distance dependence in Eq.~\eqref{eq:spectra} is given by its average value.
In this regime, the flavor transition probabilities are give by
\begin{equation}\label{avg_osc}
P(\nu_\alpha\rightarrow\nu_\beta) = \sum_{i}^{3}\left|U_{\alpha i}\right|^{2}\left|U_{\beta i}\right|^2 \approx
\left[\begin{array}{ccc}
0.55 & 0.18 & 0.27	\\
0.18 & 0.44 & 0.38	\\
0.27 & 0.38 & 0.35 
\end{array}\right],
\end{equation}
where parameters are obtained by using the best-fit values from NuFit-4.1 with normal ordering.

\subsection{Solar Flux Propagation}

Neutrinos produced in the solar center must travel through solar matter, vacuum, Earth's atmosphere, and the Earth itself to get the detector. 
In this article, we use the standard solar model given in~\cite{Vinyoles:2016djt} to propagate neutrinos from center of the Sun to the surface of the Sun.
\nusquids{} accounts for matter effects in solar matter in this process.
Assuming the DM is at rest relative to the Sun, the neutrino flux will be emitted isotropically from the center of the Sun.

To compute the expected flux at the detector, one must consider both the detector position and the Earth's position relative to the Sun. 
We can compute the time-dependent flux which changes with the solar zenith angle, as well as the time-integrated flux which depends on the time window.

\subsection{Earth Flux Propagation}

For DM clustered in the center of the Earth, neutrinos propagate from the Earth center to the detector.
The Earth density and composition is parameterized by the Preliminary Reference Earth Model (PREM)~\cite{Dziewonski:1981xy}.
Since the detector is in Earth, there is no time-dependent information of the flux; namely we assume that there is no relative motion between the DM distribution and the detector.
As discussed above, we can use \nusquids{} to propagate neutrinos from the center of the Earth to the location we are interested in. 

\section{Secluded Dark Matter\label{sec:secluded}}

Now we turn our focus to an alternative scenario in which DM particles are assumed to be {\em secluded} from the SM~\cite{Pospelov:2007mp}.
Secluded DM interacts with the SM through a dark-sector mediator while direct couplings to the SM are greatly suppressed.
\begin{figure}[b!]
  \centering
  \includegraphics[width=0.85\textwidth]{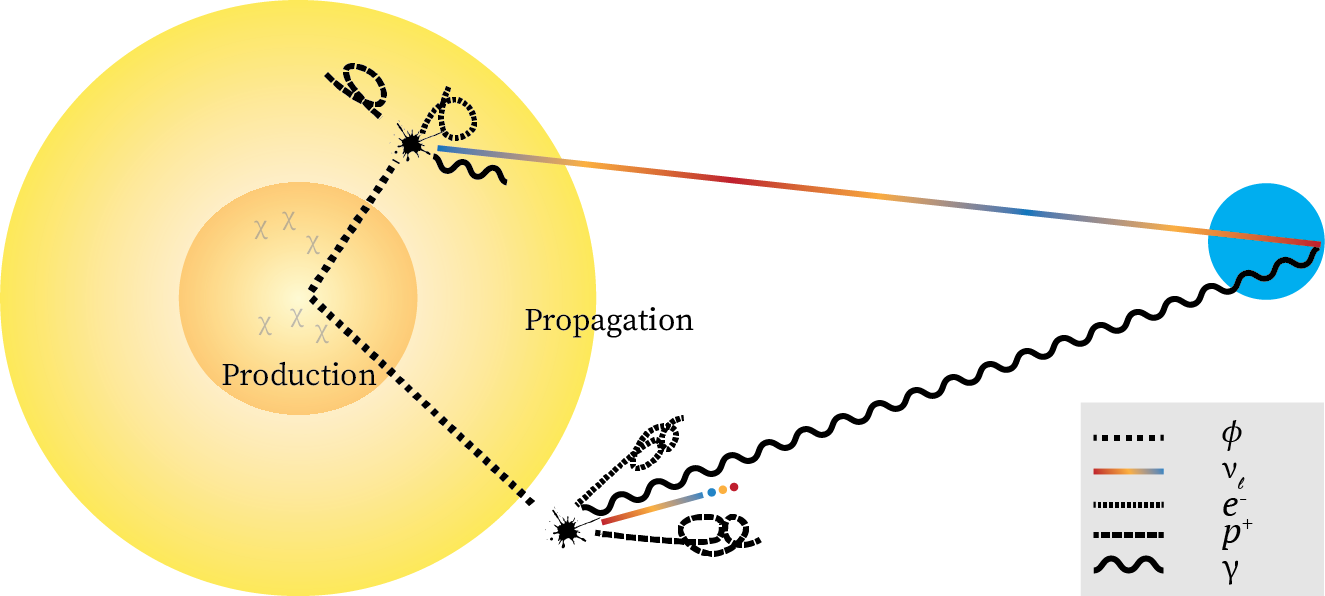}
  \caption{\textbf{\textit{Illustration of the secluded DM scenario.}}
  In this scenario the DM is secluded from the SM and predominantly communicates with it via a long-lived annihilation or decay mediator, $\phi$.
  The dotted line represents this mediator.
  If the average decay length is less than the radius of the source, then as in the standard case, only neutrinos will escape.
  On the other hand, if this length is longer than the radius of the source, then neutrinos, $\gamma$-rays, $e^{+}e^{-}$ and $p\bar{p}$ can escape the source.
  }
  \label{fig:secluded_graphic}
\end{figure}

In this model, DM annihilates to a pair of long-lived mediators, $\phi$, which then decay to SM particles; these SM particles will then decay to stable messenger particles as discussed previously.
A graphic illustrating the secluded DM scenario is shown in Fig.~\ref{fig:secluded_graphic}. 
Signatures of secluded DM from the Sun or Earth have been studied in~\cite{Pospelov:2008jd,Meade:2009mu,Schuster:2009au,Bell:2011sn,Ardid:2017lry,Leane:2017vag}.
This model includes two additional parameters besides $m_{\chi}$: $m_\phi$ and $\lambda_\phi$, the mass and decay length of the mediator.
The introduction of these new parameters affects the flux calculation in a number of ways.
For example, the ratio of $m_{\phi}$ and $m_{\chi}$ affects the angular and energy distributions of the spectrum, since at lower values of this ratio, the mediator will be highly boosted, creating a harder, more collimated spectrum.
Furthermore, $\lambda_{\phi}$ plays an important role in dense environments because as the decay length grows, the attenuation of final products decreases.
If $\lambda_{\phi}$ is larger than the radius of the neutrino-opaque environment, \textit{e.g.} $\lambda_{\phi}>r_{\odot}$ in the Sun, one can additionally look for photons, protons, electrons, and positrons as a DM signature.
In this case, the usefulness of $e^{-}$ and $p$ as messengers is case dependent since interplanetary magnetic fields can bend their trajectories such that they do not point back to their source; however, an excess of high-energy positrons or anti-protons can be a smoking-gun signature for DM~\cite{Cholis:2008qq, PhysRevLett.110.141102}.
Additionally, the introduction of $\lambda_{\phi}$ changes the production region from the point-like production region considered previously to an extended region, with the probability of SM particles being produced at a distance $r$ from the center of the source given by
\begin{equation}
    \frac{dP}{dr}(r) = \frac{e^{-r/\lambda_{\phi}}}{\lambda_{\phi}}.
    \label{eq:decay}
\end{equation}
\begin{figure}[h!]
  \centering
  \includegraphics[width=0.85\textwidth]{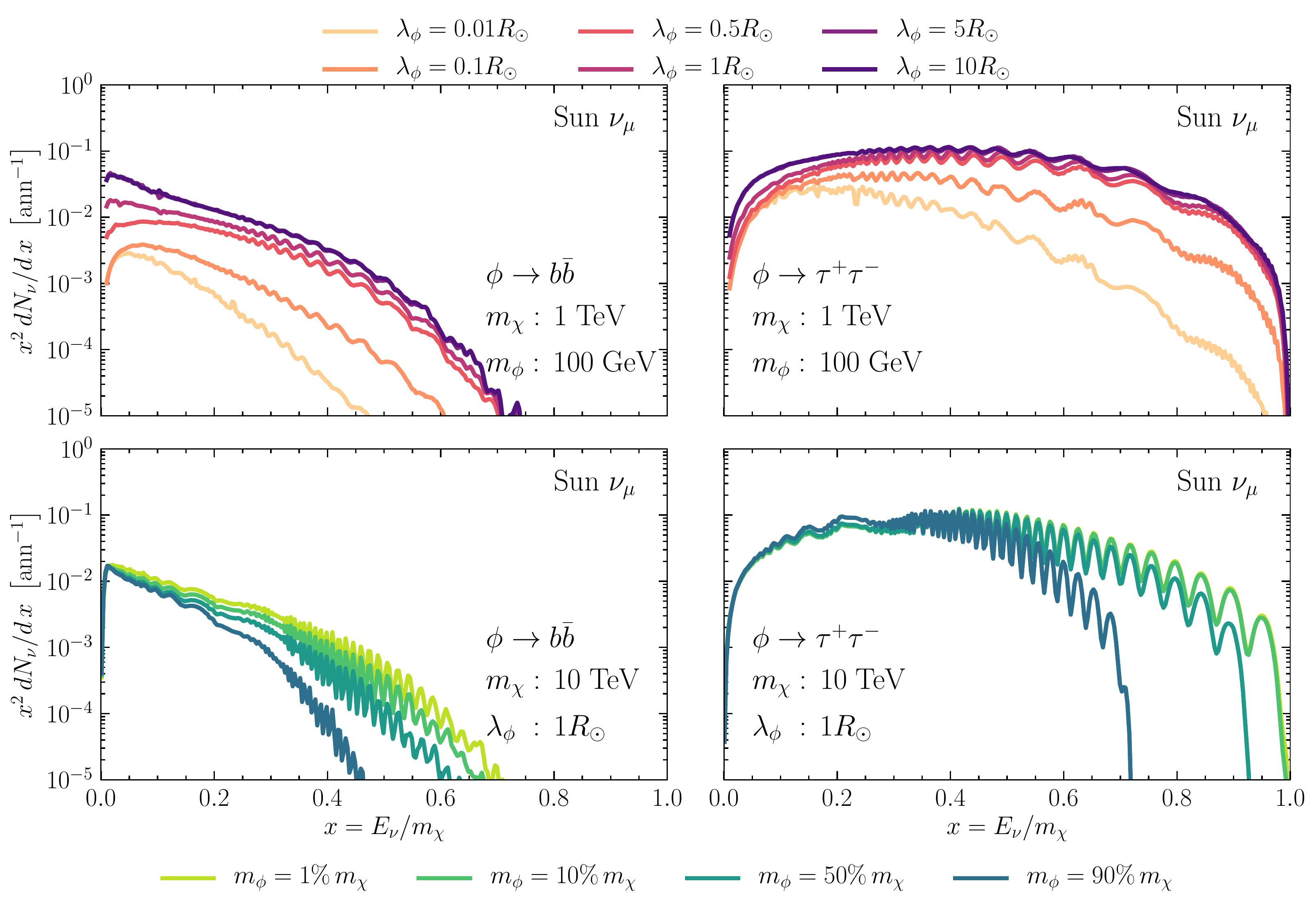}
  \caption{\textbf{\textit{Solar secluded DM neutrino fluxes at 1~AU.}}
  In the left panels we show the muon neutrino spectrum for the $b\bar b$ channel and in the right panels the $\tau^+\tau^-$ channel.
  The top plots are meant to illustrate the dependence on the mediator decay length; for this we set the mediator mass to be 100~GeV and the DM mass 1~TeV.
  The different lines show different mediator decay lengths in units of the solar radius; where they are color-sorted from shortest to longest as they go from lighter to darker.
  The bottom plots are meant to illustrate the dependence on the mediator mass, for this we now set the decay length to be one solar radius and the DM mass to be 10~TeV.
  The different lines show different mediator mass, which we show as a fraction of the DM mass; color-sorted from lightest to heaviest as they go from lighter to darker.
  }
\label{fig:secluded_sun}
\end{figure}

In order to compute the neutrino spectrum at production, we make use of the previously discussed algorithm with some additional steps that we discuss in the following lines.
First, we sample over the production region given by Eq.~\eqref{eq:decay}.
Secondly, we allow the DM to decay to the mediator with an initial energy of $m_\chi$/2.
We assume that the mediator does not interact on the way from the point of production to the point of decay.
Finally, we allow the mediator to decay and collect the energies of all neutrinos produced in the same manner as in the standard case.
In this simulation, we include interactions if the decay happens inside the Sun or Earth as discussed in Sec.~\ref{sec:production}, but with only the EW corrections included in \pythia{}; implementation of the BRW calculation is beyond the scope of this work and will be included in a future version of \charon{}.

\begin{figure}[h!]
  \centering
  \includegraphics[width=0.85\textwidth]{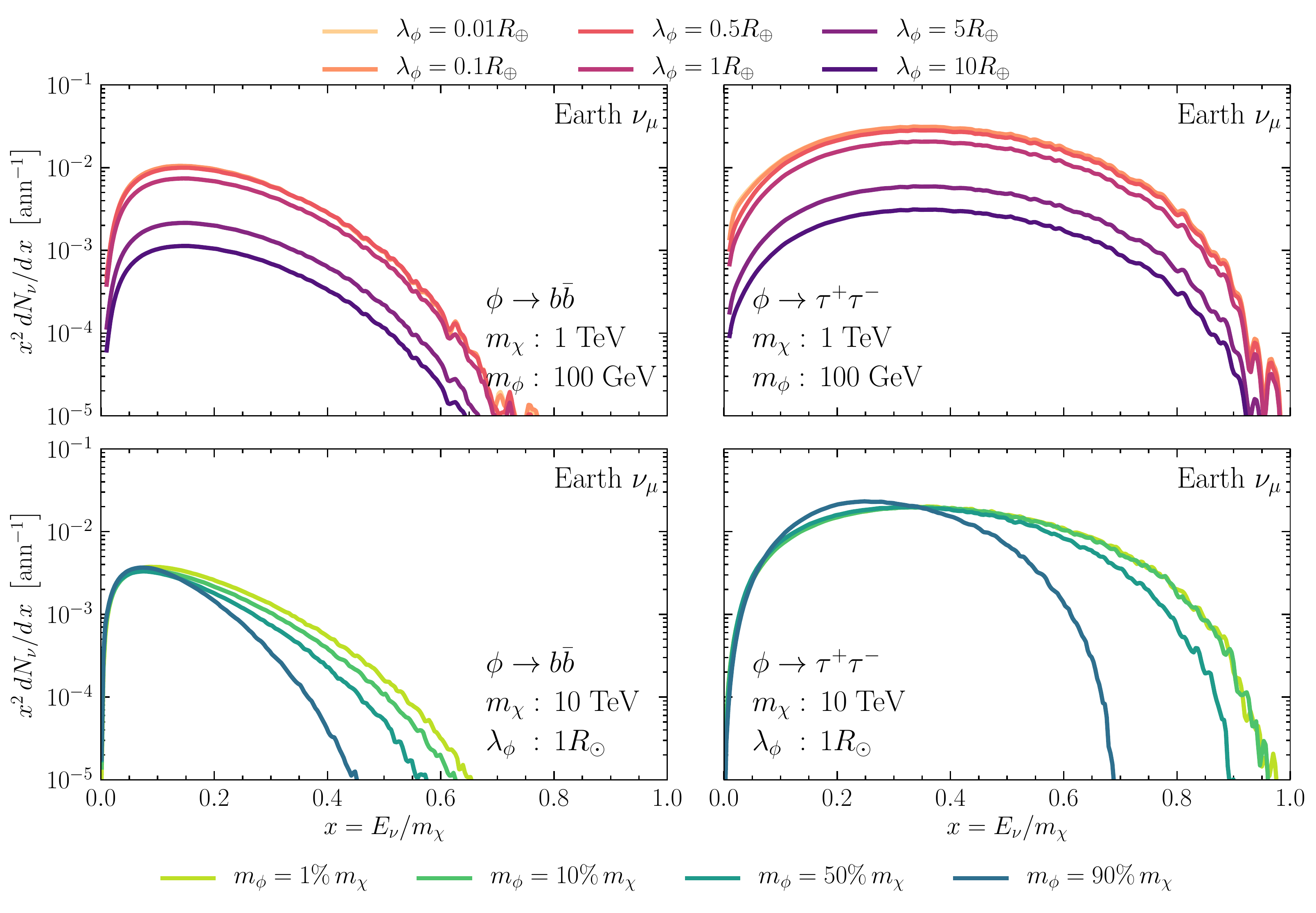}
  \caption{\textbf{\textit{Secluded DM neutrino fluxes from the Earth core at the surface.}}
  The layout and format of lines are the same as in Fig.~\ref{fig:secluded_sun}.}
  \label{fig:secluded_earth}
\end{figure}

After obtaining the flux throughout the production region, we propagate it to the detector following the process discussed in Sec.~\ref{sec:propagate}.
In this process, we use the approximation that the final neutrinos are collimated and move in the same direction as the mediator; therefore, we only consider fluxes along the line of sight from the detector to the WIMP decay location.
This approximation is valid as long as the mass of the mediator is much smaller than that of the DM; see~\cite{Niblaeus:2019gjk} for a detailed comparison using a three-dimensional simulation.
The neutrino yields for secluded DM in the Sun and Earth for a number of choices of $\lambda_{\phi}$ and $m_{\phi}/m_{\chi}$ are shown after propagation to the Earth's surface in Figs.~\ref{fig:secluded_sun} and~\ref{fig:secluded_earth}, respectively.

If $\lambda_{\phi}$ is small enough that the production region is smaller than the resolution of the detector, the flux can be treated as coming from a point source, and propagation can proceed accordingly; however, if the size of the production region is large than the detector resolution, one must treat this as an extended emission.
This requires computing a $J$-factor along a particular line of sight as described in Sec.~\ref{sec:detection}.
This extended source treatment is particularly relevant for the case of secluded DM in the Earth since the distance between the production region and detector is much shorter than in the other cases.

\section{Comparison to Other Calculations and Sources of Uncertainty\label{sec:compare}}

In this section we compare our calculation to other results in the literature and estimate the impact of different sources of uncertainties on the final neutrino flux.
For a discussion of QCD uncertainties on particle spectra from showering and hadronisation, see \cite{Amoroso:2018qga}

\subsection{Comparison to Other Calculations}

Two broadly used calculations of the flux of neutrinos from DM annihilation or decay have been used in the literature: PPPC and \wimpsim{}.
Here, we briefly comment on the differences and similarities between our result and these calculations in the three sources of neutrinos from DM discussed in this work.
\\
%

\begin{figure}
  \centering
  \includegraphics[width=\textwidth]{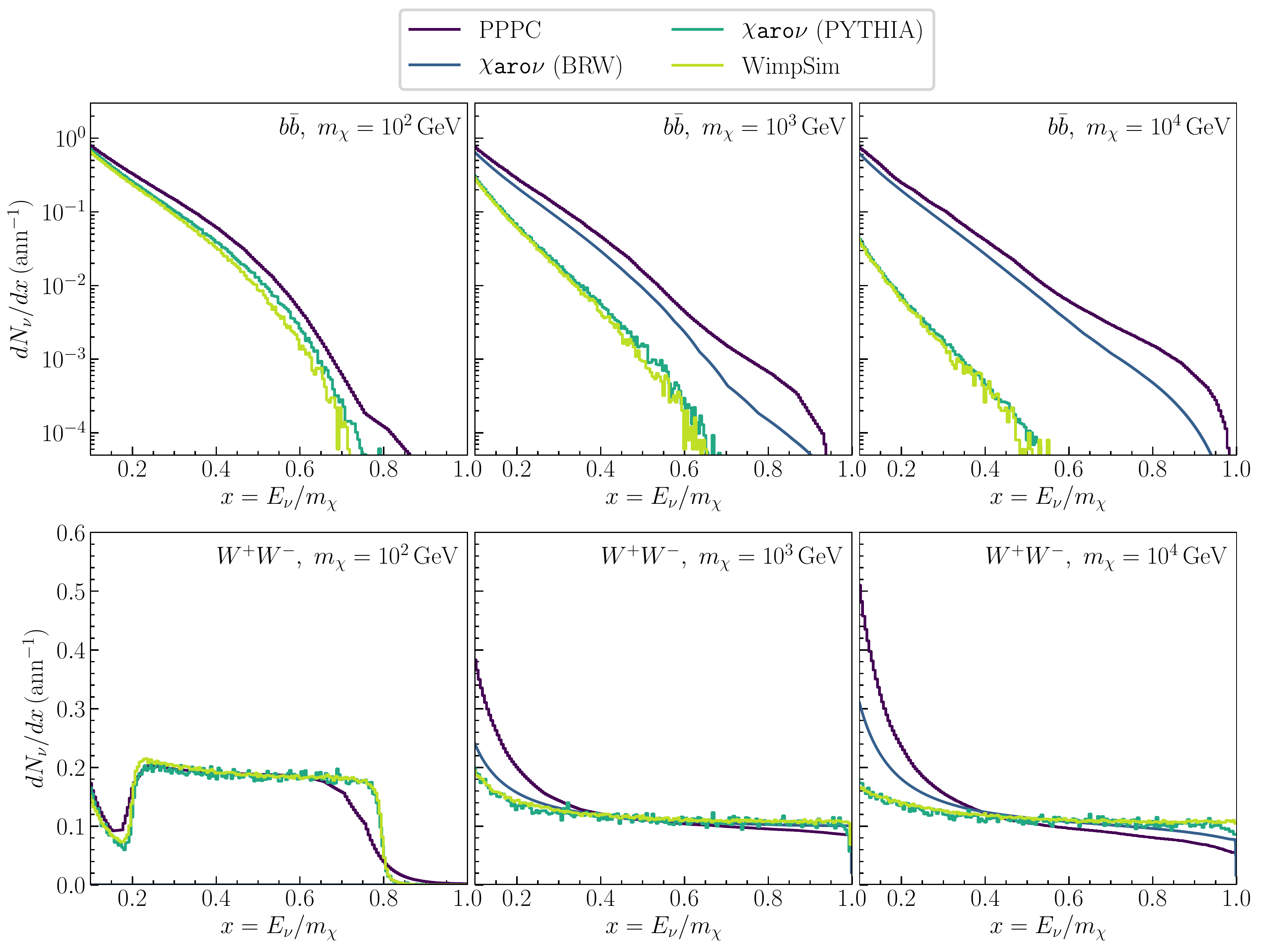}
  \caption{\textbf{\textit{Comparison of $\nu_{\mu}$-yield using four different signal generators for DM at the Sun center.}}
  The major contribution to differences between the lines is that a more complete treatment of the EW correction has been implemented in PPPC and \protect\charon{} (BRW).
  As expected, the magnitude of this difference grows as the mass of the DM increases.
  When comparing the \pythia{}-based calculations, the $b\bar b$ channel in \protect\charon{} is slightly harder than \wimpsim{} which is consistent with the result from~\cite{Cirelli:2005gh}.
  The BRW calculation does not extend to masses below 500~GeV and so it is absent from the first column.}
  \label{fig:gen_comp}
\end{figure}

\begin{figure}
  \centering
  \includegraphics[width=\textwidth]{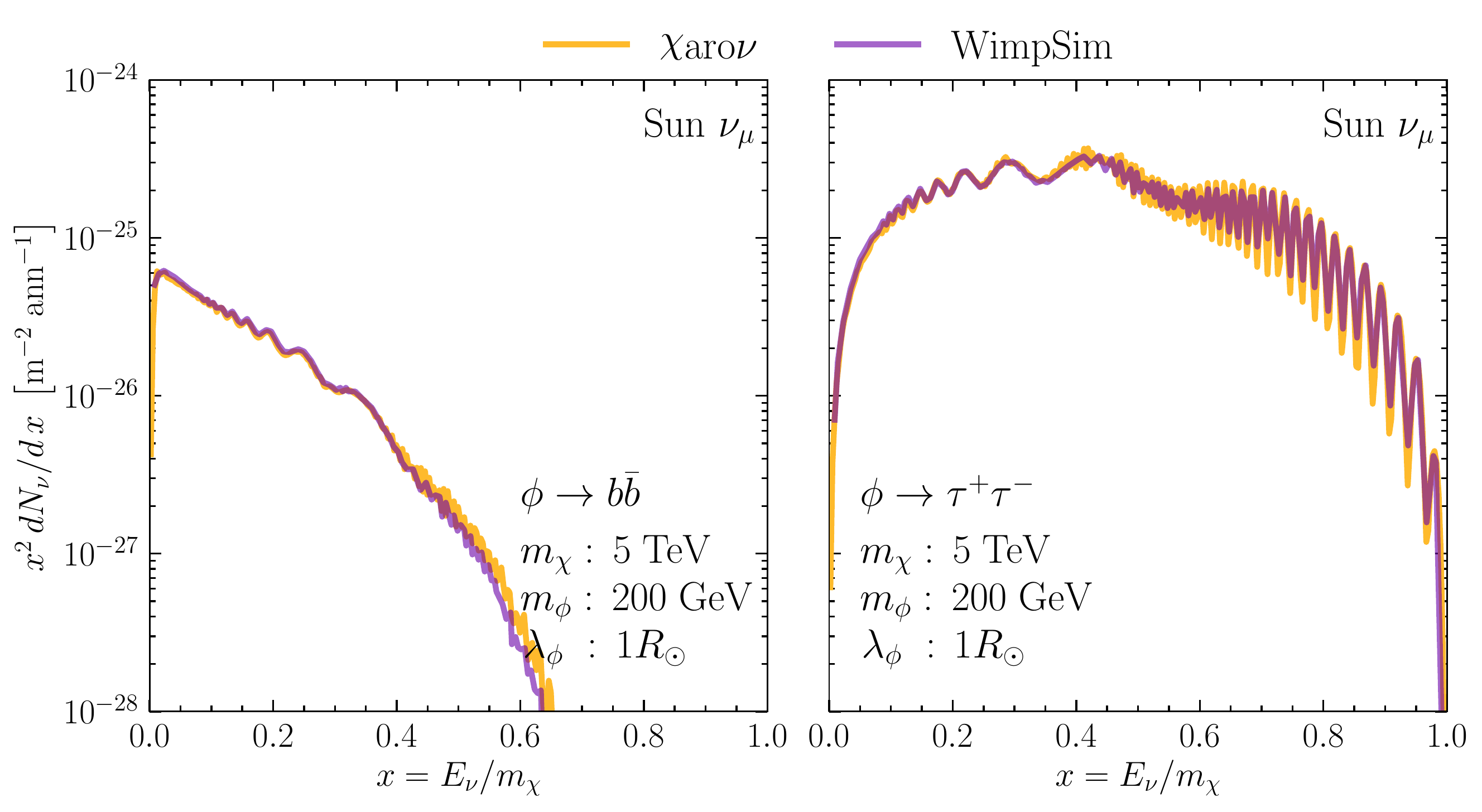}
  \caption{\textbf{\textit{Comparison of \protect\charon{} and \protect\wimpsim{} for the solar secluded case.}} 
  The \protect\wimpsim{} fluxes are from~\cite{Niblaeus:2019gjk} and oscillation parameters have been set to be the same values.
  The two fluxes are in good agreement.
  The tail of $b\bar{b}$ channel of \protect\charon{} is slightly harder than of \wimpsim{}, in agreement with our observed difference between these two calculations in the standard scenario.
  The difference of the oscillating amplitude is from interpolation which is affected by binning.}
  \label{fig:gen_comp_secluded}
\end{figure}

\noindent First for the standard DM annihilation or decay scenario we have:
\begin{itemize}
    \item \textbf{\textit{Galactic Center and Halo:}} \texttt{DarkSUSY}~\cite{Gondolo:2004sc,Bringmann:2018lay}, PPPC, or direct \pythia{} calculation, has been used in results predominantely from DM annihilation from the Galactic Halo, see \textit{e.g.} results by IceCube~\cite{Aartsen:2017ulx,Aartsen:2020tdl}, ANTARES~\cite{Albert:2016emp,Aartsen:2020tdl}, and Super-Kamiokande~\cite{Abe:2020sbr}.
    Our calculation using \pythia{}\texttt{8.2} matches well with previous calculations with \pythia{}\texttt{6}.
    Due to the fact that \pythia{} only partially includes EW interactions both predicted fluxes are smaller than those predicted by PPPC, \textit{e.g.} the PPPC calculation gives spectra which are $\sim$ 20 times larger for DM annihilation to $e^{-}e^{+}$ with $m_{\chi}=1~\rm{TeV}$.
    When we incorporate EW corrections, by coupling our calculation with BRW~\cite{Bauer:2020jay}, our results yield a flux greater than the \pythia{} versions, but less than the PPPC calculation by an average factor of 1.8, for the same parameters mentioned above, due to different approaches discussed in Sec.~\ref{sec:production}.

    \item \textbf{\textit{Sun:}} Though PPPC provides a calculation of DM annihilation from the Sun, most experimental results use \wimpsim{}.
    These calculations differ from each other as well as from \charon{} in several meaningful ways, both in their treatment of fluxes at production, and in neutrino transport.
    
    With respect to production, both PPPC and \charon{} include a treatment of the EW correction, which is not implemented in \wimpsim{}.
    Furthermore, PPPC handles interactions of stable and metastable particles using \geant{}, which tracks low-energy neutrinos produced in interactions of primary hadrons with the environment.
    These neutrinos are ignored by \wimpsim{}, and as mentioned above, \charon{} allows the inclusion of these low-energy neutrinos as an option.
    Lastly, PPPC samples energy losses of $b$ and $c$ hadrons from a distribution, whereas \charon{} and \wimpsim{} use the average energy loss for a given interaction.
    Though small, this effect is most notable notable in hadronic channels, \textit{e.g.} $b\bar{b}$~\cite{Baratella:2013fya}.
    
    With respect to propagation, both PPPC and \nusquids{} use an integro-differential equation approach, while \wimpsim{} uses a Monte Carlo based transport. 
    The Monte Carlo approach allows for event-by-event simulation, which offers several advantages.
    Specifically, it is easier to couple such simulations to detector simulations and allows tracking the position of the Sun on an event-by-event basis.
    These advantages, however, come at the cost of significantly increasing propagation time.
    Additionally, the \wimpsim{} calculation uses a neutrino cross section which uses the CTEQ6 parton distribution functions which which yields a more accurate cross section than other calculations at small-$Q^2$ values, \textit{e.g.}~\cite{CooperSarkar:2011pa} that use HERAPDF.
    Our calculation is  more similar to PPPC than to \wimpsim{}, and has the following notable differences compared to the latter: we treat DM emission to SM particles as a point-source; we propagate the neutrinos using a differential equation instead of a Monte Carlo method, we take into account the polarization of the $\tau$ when considering tau-neutrino regeneration, and, when coupled with BRW calculation~\cite{Bauer:2020jay}, have a more complete treatment of EW corrections. 
    As can be seen in Fig.~\ref{fig:gen_comp} for a WIMP of 100~GeV our calculation and \wimpsim{}'s are in agreement.
    For DM masses above the EW scale, for definiteness at $10^3$~GeV and $10^4$~GeV in the figure, our calculation that includes EW corrections is significantly larger than the \wimpsim{} and \pythia{}-only predictions. 
    In this range the effects can be as large as a factor of four for the $b\bar{b}$ channel with $m_{\chi}=1~\rm{TeV}$.
    Lastly, we find the effect of PPPC's partial implementation of EW corrections is channel dependent.
    For example, in the $b\bar{b}$ channel, PPPC tends to overpredict the flux relative to our calculation over the whole energy range, whereas in the $W^{+}W^{-}$ channel PPPC overpredicts the flux at low energies, but underpredicts it at high energies.
    \item \textbf{\textit{Earth:}} \wimpsim{} has been used in recent searches for DM annihilation from the Earth, see \textit{e.g.} results by ANTARES~\cite{Albert:2016dsy}, IceCube~\cite{Aartsen:2016fep} and Super-Kamiokande~\cite{Frankiewicz:2017trk}.
    In this case, we find that our results agree well with \wimpsim{} when using \pythia{} to generate the fluxes and are larger when seeding with the BRW~\cite{Bauer:2020jay} calculation as discussed in the Sun case, as expected. 
\end{itemize}
Second for the secluded DM scenario:
\begin{itemize}
    \item \textbf{\textit{Sun:}} This scenario was implemented in \wimpsim{} in~\cite{Niblaeus:2019gjk}.
    Beyond the differences already discussed between \wimpsim{} and \charon{}, it should be noted that our implementation of secluded DM is a one-dimensional calculation.
    In~\cite{Niblaeus:2019gjk}, a comparison with a three-dimensional implementation is performed, concluding that effect is only important when the mediator mass, $m_\phi$, is close to the DM mass, $m_\chi$.
    This causes the products of the mediator not to be collinear with the DM direction and has a 28\% effect on average on the final flux when the mediator mass is 98\% of the DM mass. 
    For dark mediator mass fractions considered in this work this effect is negligible, and our calculations are in good agreement as can be seen in Fig.~\ref{fig:gen_comp_secluded}. 
    Additionally, if the decay of the mediator is in or near the detector, a di-muon signature from the muon decay channel would be induced, creating a smoking-gun signature of DM~\cite{Meade:2009mu,Ardid:2017lry,Niblaeus:2019gjk}.
    \item \textbf{\textit{Earth:}} The secluded DM scenario has not been implemented in any previous result and is included for the first time in \charon{}.
    Considering this scenario in the Earth is interesting for several reasons.
    Long-range interactions can induce Sommerfeld-enhanced annihilation cross sections~\cite{Delaunay:2008pc, Feng:2016ijc}.
    Similar to the solar case, if the mediator decays to muons and it happens near the detector, it induces a di-muon signature from the direction of the core~\cite{Meade:2009mu}, which is complementary to the neutrino flux predicted here.
\end{itemize}

\subsection{Effects of the Neutrino Oscillation Parameters}

To estimate the effect that uncertainties on the neutrino oscillation parameters have on the neutrino yields, we construct a two-sided Gaussian for each parameter centered on the best-fit point from NuFit-4.1~\cite{nufit:2018abc}, whose standard deviations are the $\pm 1\sigma$ values from the same work.
We then sample from each Gaussian independently, and run \charon{} using the resulting parameters.
We repeat this procedure to create a distribution of fluxes, and create the highest-density posterior interval at given credible levels.

\begin{figure}
  \centering
  \includegraphics[width=1\textwidth]{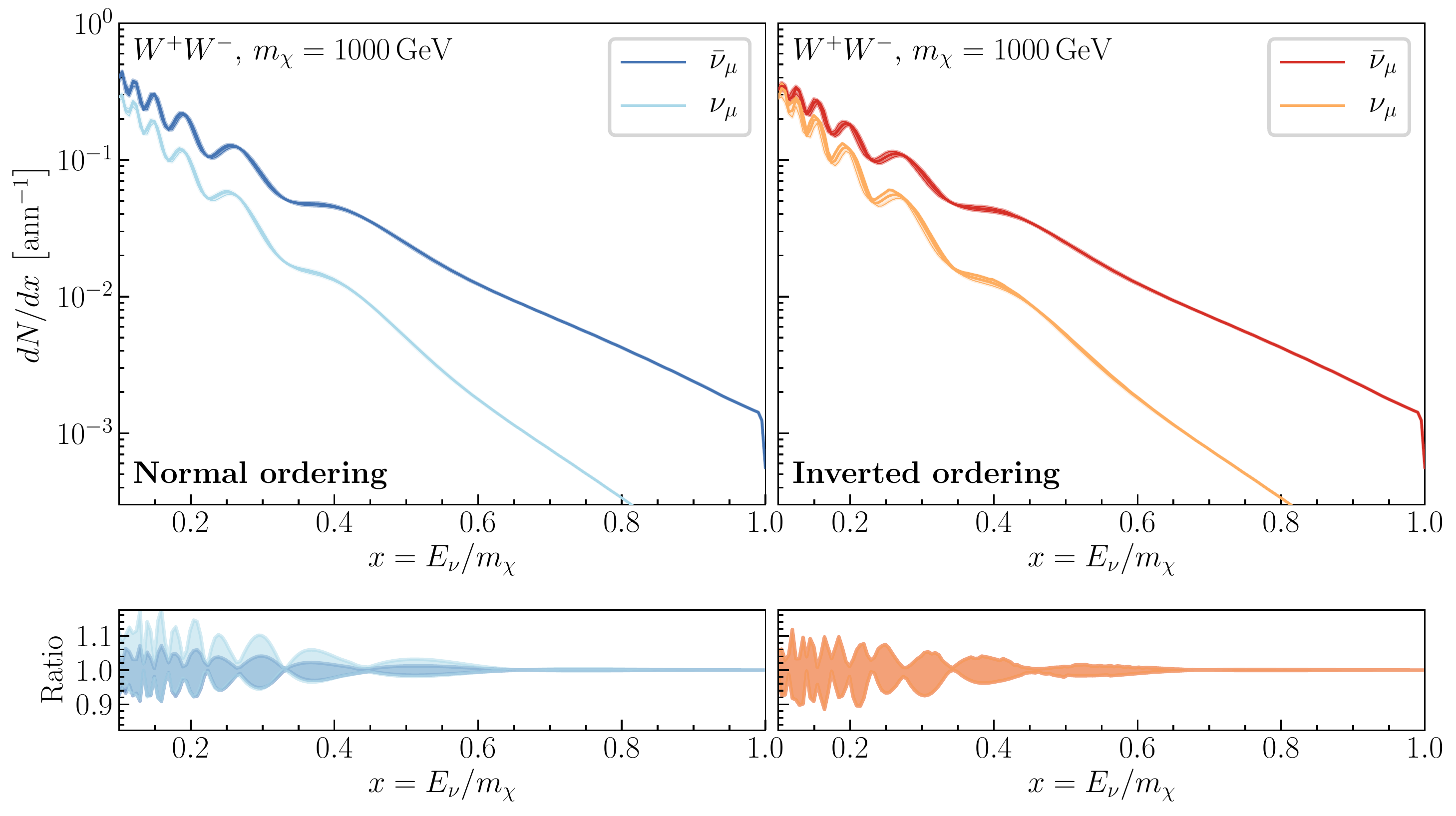}
  \caption{\textbf{\textit{Effects of oscillation parameters and neutrino ordering.}}
  We vary each oscillation parameter according to the procedure described in the text.
  In both panels we assume annihilation to $W^+W^-$ with $m_{\chi}=$1~TeV; on the left plot we assume normal neutrino ordering and on the right one inverted ordering.
  The envelopes shows the 2$\sigma$ uncertainty regions.
}
  \label{fig:oscillation_impact}
\end{figure}

The results of this procedure at the 2$\sigma$ level are shown for $\chi\chi\rightarrow W^{+}W^{-}$ and $m_{\chi}=10^{3}$~GeV in Fig.~\ref{fig:oscillation_impact}, where this is performed separately for normal and inverted ordering.
The uncertainty 95\% containment region gives an uncertainty of up to 15\% at energies below $\sim$200~GeV, with these differences shrinking to a few percent for higher energy neutrinos.

\subsection{Effects of the Neutrino Cross Section}

\begin{figure}
  \centering
  \includegraphics[width=1\textwidth]{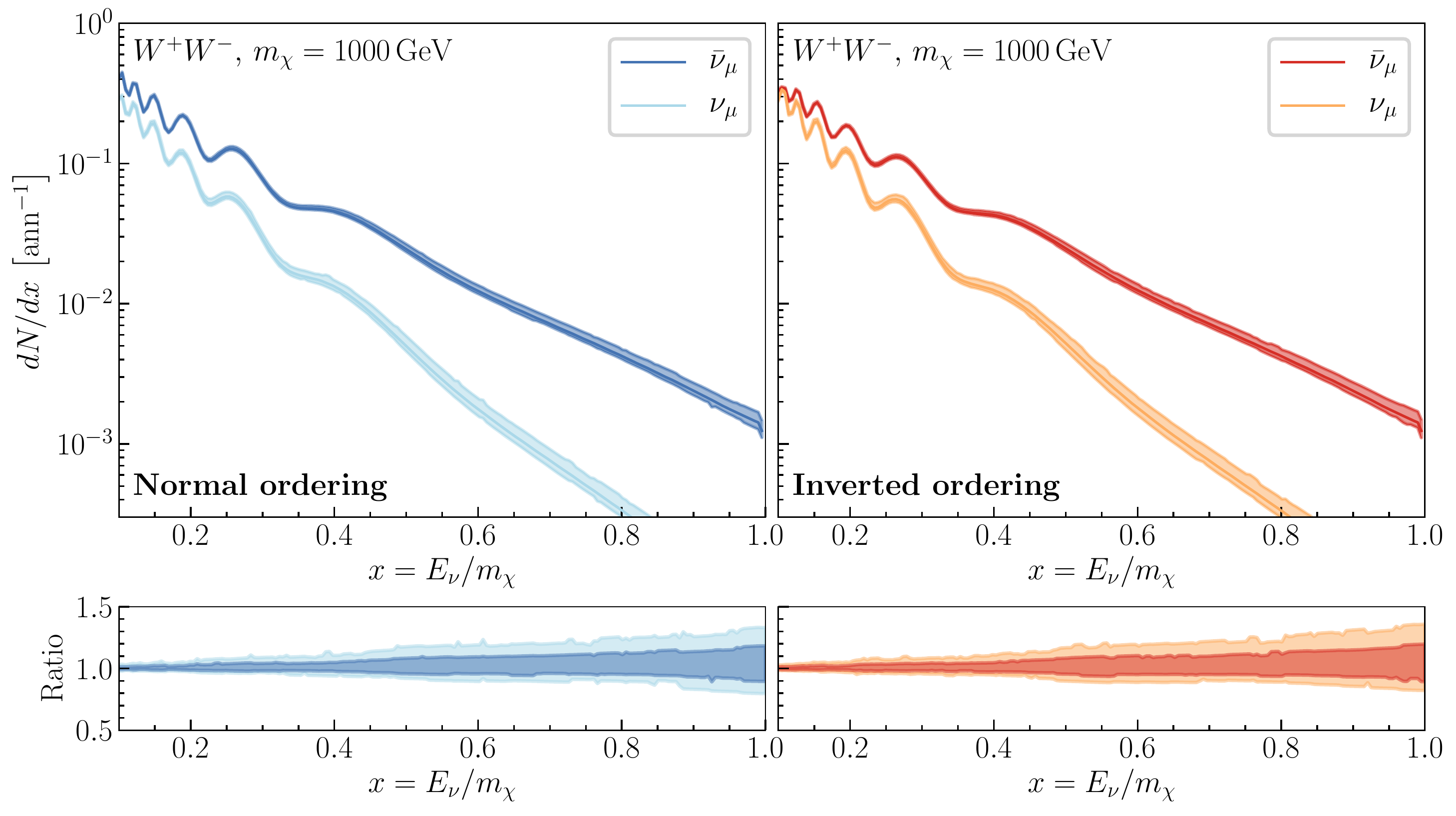}
  \caption{\textbf{\textit{Effects of the neutrino cross section uncertainties.}}
  We vary cross sections according to the procedure described in the text.
  In both panels we assume annihilation to $W^+W^-$ with $m_{\chi}=$1~TeV; on the left plot we assume normal neutrino ordering and on the right one inverted ordering.
  We use the CSMS cross section model~\cite{CooperSarkar:2011pa}, and show the 2$\sigma$ using the uncertainties from this work.
  }
  \label{fig:crosssection_impact}
\end{figure}

To estimate the effect that cross section uncertainties have on the neutrino flux, we follow a procedure similar to the previously discussed procedure.
In this case, however, we construct a two-sided Gaussian centered around one, and the sigma values are the percent uncertainty given in~\cite{CooperSarkar:2011pa}, using the central value from this same work, which we assume to be comparable to the \texttt{NuSigma} cross sections.
We then sample from these distributions at each energy, and rescale the default \nusquids{} cross section by the drawn value.
We repeat this procedure to create a distribution of fluxes, and create the highest-density posterior interval at given credible levels.
The result of this procedure is shown in~\ref{fig:crosssection_impact}.

The effect of cross section uncertainties is quite small at low energies as the opacity is small at these energies, but it grows with energy.
At the highest energies, the error on the flux reaches $\sim$50\% for $\nu_{\mu}$ and $\sim$25\% for $\bar{\nu}_{\mu}$. 
The reason for the increasing uncertainty can be understood from the fact that the survival probability, $p_{\rm sur.}$, is given by $\exp\left(-n \sigma(E_\nu)\right)$, where $n$ is total column density traversed and $\sigma(E_\nu)$ the total cross section at a given neutrino energy, $E_\nu$.
This implies that the relative uncertainty is $\delta\left[p_{\rm sur.}\right]/p_{\rm sur.} \propto \delta \left[\sigma (E_\nu)\right]$, which increases with energy.

\section{Conclusion\label{sec:conclusion}}

In this work we have computed the flux of neutrinos from DM decay and annihilation both in the standard WIMP paradigm as well as in the secluded DM scenario.
Our calculation includes several updates over previous results such as new EW corrections for neutrino production using the BRW calculation~\cite{Bauer:2020jay}, as well as a new propagation tool that is flexible and allows one to vary parameters, such as the neutrino oscillation parameters and neutrino-nucleon cross sections.
Additionally, we include the possibility of the secluded DM scenario in the Sun and Earth, the latter of which is a new result.
In our new result we find that, when including EW corrections consistently, the flux of neutrinos from DM annihilation or decay for DM masses above the EW scale is increased by a factor of $\sim$~2 for 500~GeV and $\sim$~100 for 10~TeV for $b\bar b$ channel on average.
Since the effect of this, in the range where neutrinos are able to exit their production environment, is to approximately rescale the flux, current constraints on DM annihilation from the Sun and Earth are underestimated by approximately this same factor.

\acknowledgments

We thank Marco Cirelli, Joakim Edsj\"o and Francis Halzen for useful discussions.
We also thank Christian Bauer, Nicholas Rodd, and Bryan Webber for a fruitful collaboration.
We thank Peter Denton, Alejando Diaz, Matheus Hostert, Zach Krebs, Ibrahim Safa, and Carsten Rott for comments on the draft.
We would like to give special thanks to Sergio Palomares-Ruiz and Aaron Vincent for their detailed comments that have greatly improved this work.
CAA was supported by NSF grant PHY-1912764.
AK acknowledges the support from the IGC Postdoctoral Award.
QL and JL are supported by NSF under grants PLR-1600823 and PHY-1607644 and by the University of Wisconsin Research Council with funds granted by the Wisconsin Alumni Research Foundation.

\bibliographystyle{JHEP-2}
\bibliography{bibfile}

\end{document}